\documentclass[twocolumn,showpacs,preprintnumbers,amsmath,amssymb,superscriptaddress,groupedaddress,prmater]{revtex4}

\usepackage{epsfig}                                                            
\usepackage{graphicx}
\usepackage{dcolumn}
\usepackage{color}
\usepackage{bm}
\usepackage{subfigure}
\newcommand{\be}{\begin{equation}} \newcommand{\ee}{\end{equation}}
\newcommand{\bea}{\begin{eqnarray}} \newcommand{\eea}{\end{eqnarray}}

\begin{document}

\title{\bf Room temperature surface multiferroicity in Y$_2$NiMnO$_6$ nanorods}

\author {Shubhankar Mishra} \affiliation {School of Materials Science and Nanotechnology, Jadavpur University, Kolkata 700032, India}
\author {Amritendu Roy} \affiliation {School of Minerals, Metallurgical and Materials Engineering, Indian Institute of Technology, Bhubaneswar 752050, India} 
\author {Aditi Sahoo} \affiliation {Advanced Materials and Chemical Characterization Division, CSIR-Central Glass and Ceramic Research Institute, Kolkata 700032, India; Present Address: Department of Physics, Indian Institute of Technology, Gandhinagar 382355, India}
\author {Biswarup Satpati} \affiliation {Saha Institute of Nuclear Physics, A CI of Homi Bhabha National Institute, 1/AF Salt Lake, Kolkata 700064, India}
\author {Anirban Roychowdhury} \affiliation {Department of Physics, Krishnath College, Berhampore 742101, West Bengal, India} 
\author {P.K. Mohanty} \affiliation {Department of Physical Sciences, IISER, Kolkata, Mohanpur, West Bengal 741246, India} 
\author {Chandan Kumar Ghosh} \email {chandu_ju@yahoo.co.in} \affiliation {School of Materials Science and Nanotechnology, Jadavpur University, Kolkata 700032, India}
\author {Dipten Bhattacharya} \email {dipten@cgcri.res.in} \affiliation {Advanced Materials and Chemical Characterization Division, CSIR-Central Glass and Ceramic Research Institute, Kolkata 700032, India}

\date{\today}

\begin{abstract}
We report observation of surface-defect-induced room temperature multiferroicity - surface ferromagnetism ($M_S$ at 50 kOe $\sim$0.005 emu/g), ferroelectricity ($P_R$ $\sim$2 nC/cm$^2$), and significantly large magnetoelectric coupling (decrease in $P_R$ by $\sim$80\% under $\sim$15 kOe field) - in nanorods (diameter $\sim$100 nm) of double perovskite Y$_2$NiMnO$_6$ compound. In bulk form, this system exhibits multiferroicity only below its magnetic transition temperature $T_N$ $\approx$ 70 K. On the other hand, the oxygen vacancies, formed at the surface region (thickness $\sim$10 nm) of the nanorods, yield long-range magnetic order as well as ferroelectricity via Dzyloshinskii-Moriya exchange coupling interactions with strong Rashba spin-orbit coupling. Sharp drop in $P_R$ under magnetic field indicates strong cross-coupling between magnetism and ferroelectricity as well. Observation of room temperature magnetoelectric coupling in nanoscale for a compound which, in bulk form, exhibits multiferroicity only below 70 K underscores an alternative pathway for inducing magnetoelectric multiferroicity via surface defects and, thus, in line with magnetoelectric property observed, for example, in domain walls or boundaries or interfaces of heteroepitaxially grown thin films which do not exhibit such features in their bulk.
\end{abstract}
\pacs{75.70.Cn, 75.75.-c}
\maketitle

\section{Introduction}

\begin{figure}[ht!]
\begin{center}
   \subfigure[]{\includegraphics[scale=0.30]{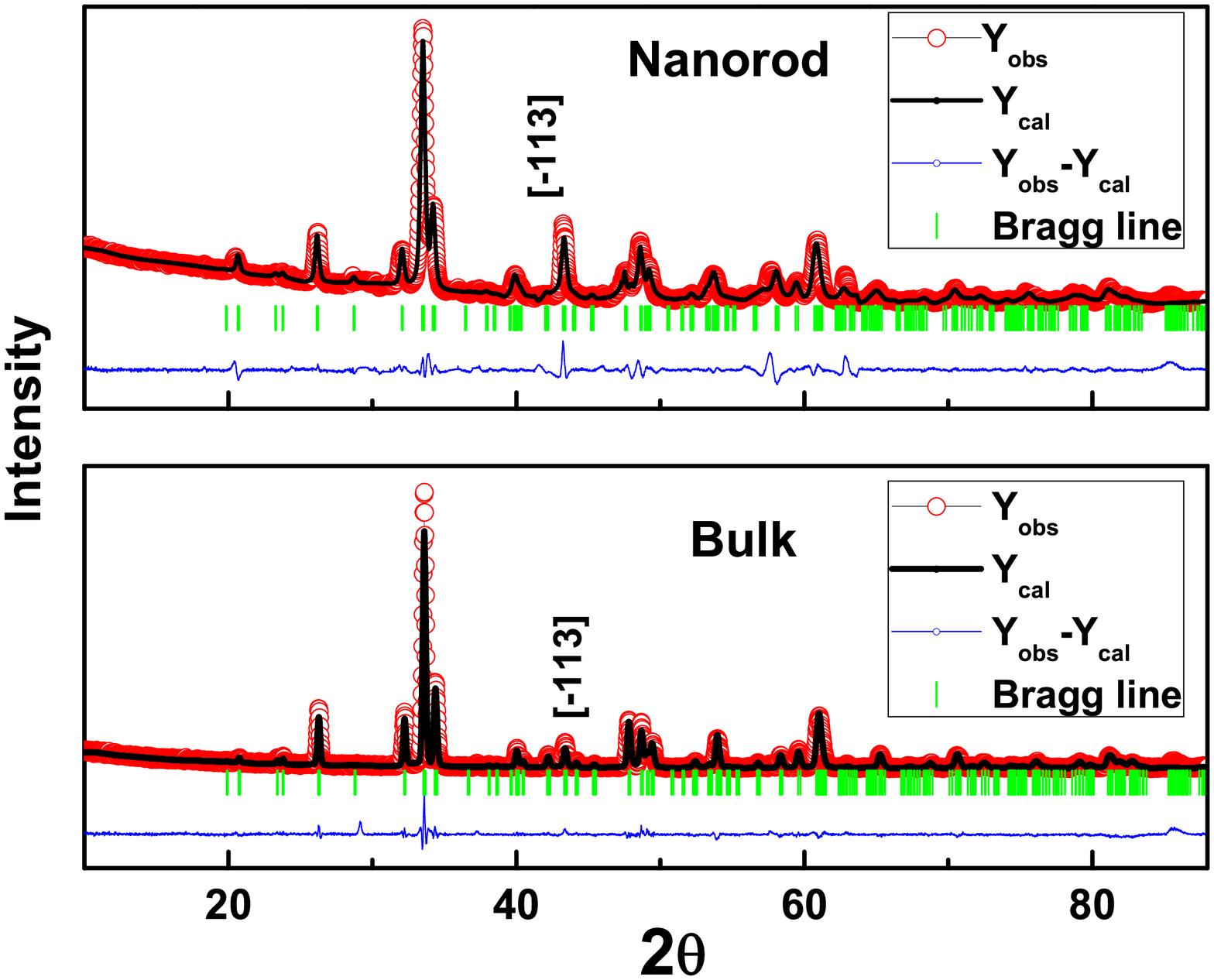}} 
   \subfigure[]{\includegraphics[scale=0.20]{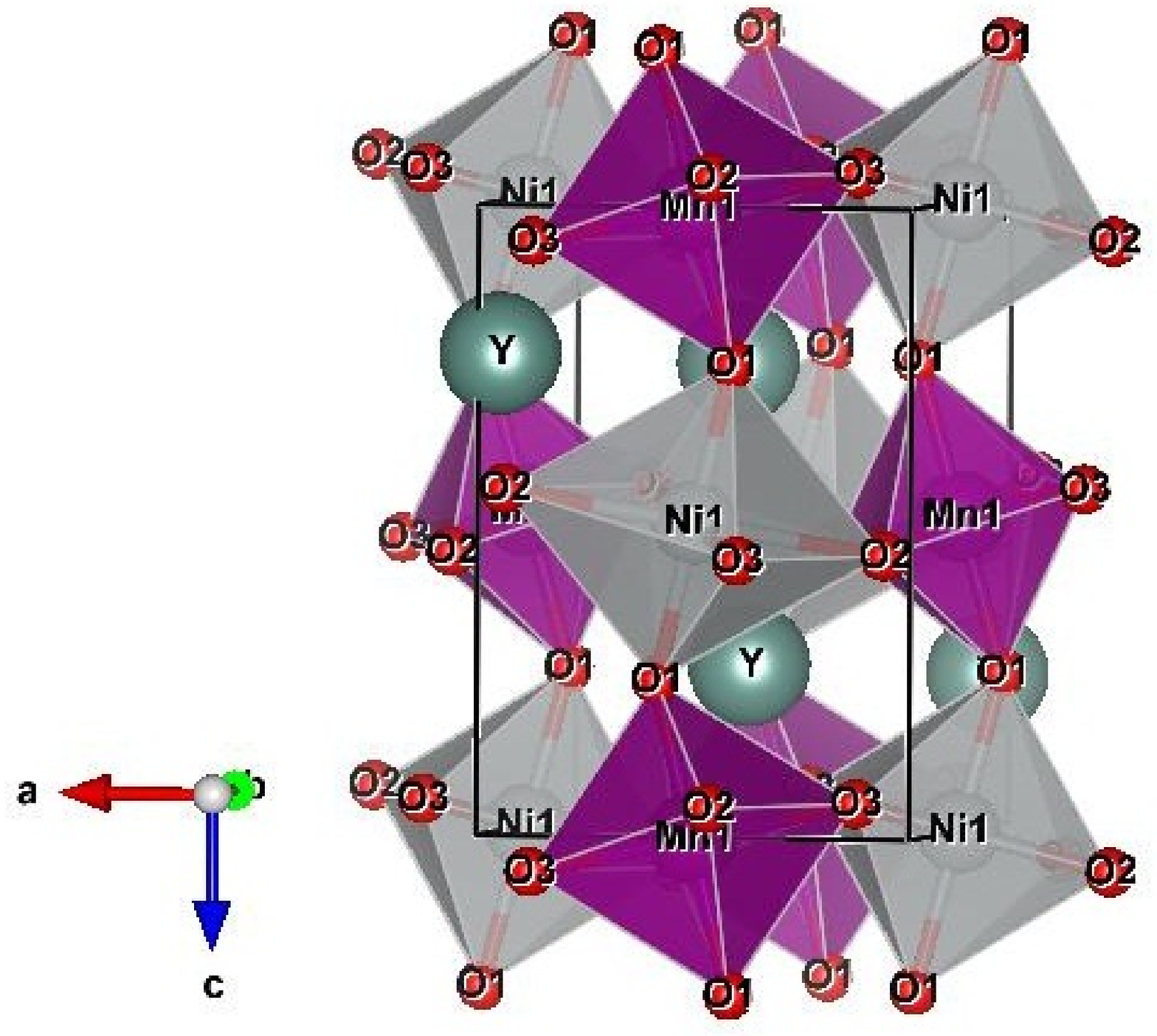}}
   \end{center}
\caption{(a) The room temperature x-ray diffraction patterns for bulk and nanoscale samples and their refinement by FullProf; (b) crystallographic structure of Y$_2$NiMnO$_6$. }
\end{figure}

Alongside comprehensive investigation of multiferroicity in single-phase Type-I and Type-II systems and composites, in recent time, an alternative paradigm is emerging where spontaneously formed (such as domain wall or boundaries) and/or artificially fabricated (such as nanosized and heteroepitaxial architectures) surface/interface regions are found to be hosting the magnetic and ferroelectric orders and offering the test bed for inducing coupling among the order parameters \cite{Fiebig,Zutic,Spaldin}. For example, ferroelastic domain walls in SrTiO$_3$, below its structural phase transition (from cubic to tetragonal) at $\sim$105 K, host ferroelectric order and, therefore, respond to electrical tuning and, in turn, change the magnetic domain structure in La$_{\frac{1}{2}}$Sr$_{\frac{1}{2}}$MnO$_3$ deposited on the SrTiO$_3$ substrate \cite{Salje}. It has also been pointed out in a theoretical work \cite{Bellaiche} that, in orthoferrite SmFeO$_3$, magnetic domain boundaries could be polar while the domains themselves are not. Surface and interface regions in many systems have been found to be topologically protecting the magnetic and electrical vortices with often a coupling between them \cite{Mathur,Tokura,Loidl,Kezsmarki,Nahas,Wang,Goncalves,Das}. They were shown to result from extended spin-orbit coupling in lower dimension \cite{Fert}. In a recent theoretical work \cite{Betouras}, it has been claimed that a bulk and collinear ferro- or antiferromagnet supports surface ferroelectricity both in presence and absence of surface induced Dzyaloshinskii-Moriya (DM) exchange interaction. These new developments are, therefore, precipitating a new paradigm of surface/interface multiferroicity which is distinct even from multiferroicity in the composite systems where striction mediated coupling between magnetic and ferroelectric order parameters across the interfaces of constituent phases is the key feature. Given this background, a simple question naturally arises: whether in the nanosized particles as well, surface induced multiferroicity is possible (because of their enhanced surface area to volume ratio) in the absence of bulk multiferroicity. Surface ferromagnetism was already shown \cite{Sundaresan} to be ubiquitous in nanosized materials even in the absence of magnetic ions. However, it is not known whether ferroelectricity could also emerge along with surface magnetism in nanoscale. Earlier work \cite{Lu} on nanoscale multiferroic systems such as BiFeO$_3$, o-TbMnO$_3$, h-RMnO$_3$ (R = Sm, Eu, Gd, Dy, Tb) etc concentrated only on exploring the prevalence of bulk multiferroicity as a function of particle size and/or thickness of films. In this paper, we demonstrate that in nanorods of double perovskite Y$_2$NiMnO$_6$ (YNMO) compound - which in bulk form exhibits ferroelectricity due to magnetic ordering below $T_N$ $\approx$ 70 K - magnetoelectric coupling between surface ferromagnetism and surface ferroelectricity is quite significant. The oxygen vacancies at the surface induce surface ferromagnetism at room temperature while surface ferroelectricity emerges from Dzyloshinskii-Moriya (DM) exchange coupling interactions within the noncentrosymmetric surface in presence of large Rashba spin-orbit coupling. Measurement of remanent ferroelectric polarization under varying magnetic field at room temperature shows that the magnetoelectric coupling is substantial. The bulk YNMO assumes centrosymmetric $P2_1/n$ crystallographic structure at room temperature with ordering of Ni$^{2+}$/Mn$^{4+}$ ions. It exhibits reasonably strong magnetic order driven ferroelectricity ($P_S$ $\sim$ 0.15 $\mu$C/cm$^2$) below $T_N$ $\approx$ 70 K \cite{Su}. The magnetic structure turns out to be exchange striction driven collinear E-type antiferromagnetic ($\uparrow\uparrow\downarrow\downarrow$). The ferroelectricity in such systems arises from asymmetric shift of the O$^{2-}$ ions. 

\section{Experimental Details}

The YNMO nanorods were prepared by hydrothermal method. The nitrate and acetate salts such as [Y(NO$_3$)$_3$,6H$_2$O], [Ni(NO$_3$)$_3$,6H$_2$O] and C$_4$H$_6$MnO$_4$ were first dissolved in 40 ml deionized water in stoichiometric ratio (2:1:1 molar ratio) under stirring. Then 10 ml of 5M NaOH solution was added. The precipitate formed was collected and stirred for 30 min. It was later transferred to a Teflon sealed stainless steel autoclave and heated at 180$^o$C for 24h. The final product was centrifuged by distilled water and ethanol and dried at 70$^o$C for 24h. The powder obtained thus was ground by mortar and calcined at 1000$^o$C for 4h. Bulk YNMO is prepared by solid state reaction method. The bulk and nanosized YNMO were characterized by x-ray diffraction (XRD), scanning electron and transmission electron microscopy (SEM and TEM), and x-ray photoelectron spectroscopy (XPS). The XRD data were refined by FullProf. X-ray photoelectron spectroscopy measurement was carried out to determine the charge states of the ions in both bulk sample and nanorods of YNMO. Temperature dependent magnetic properties were measured by a vibrating sample magnetometer (Cryogenics 16T VSM). The magnetic force microscopy (MFM) images were captured by the LT-AFM/MFM system of NanoMagnetics Instruments using commercial Co-alloy coated MFM cantilevers. Two-pass mode has been used in raster scan with cantilever oscillating at the resonant frequency ($\sim$70 kHz) by digital phase-lock-loop control system. The oscillation amplitude varies within 10-50 nm. The forward scan records the surface topography in semi-contact mode where cantilever oscillation amplitude is used as the feedback parameter. The cantilever is then lifted by 50-150 nm from the surface to reduce the influence of short-range force and record the magnetic interaction between tip of the cantilever and the sample surface. The corresponding phase shift is recorded as the magnetic domain image. To measure the ferroelectric polarization, YNMO nanorods were deposited on Si/SiO$_2$ substrate and silver electrodes were deposited in two-probe top-top electrode configuration. For bulk pellet samples, silver electrodes were used in top-bottom configuration. The remanent ferroelectric hysteresis loops were measured by Ferroelectric Loop Tester of Radiant Inc. (Precision LC-II model).

\begin{figure*}[ht!]
\begin{center}
   \subfigure[]{\includegraphics[scale=0.18]{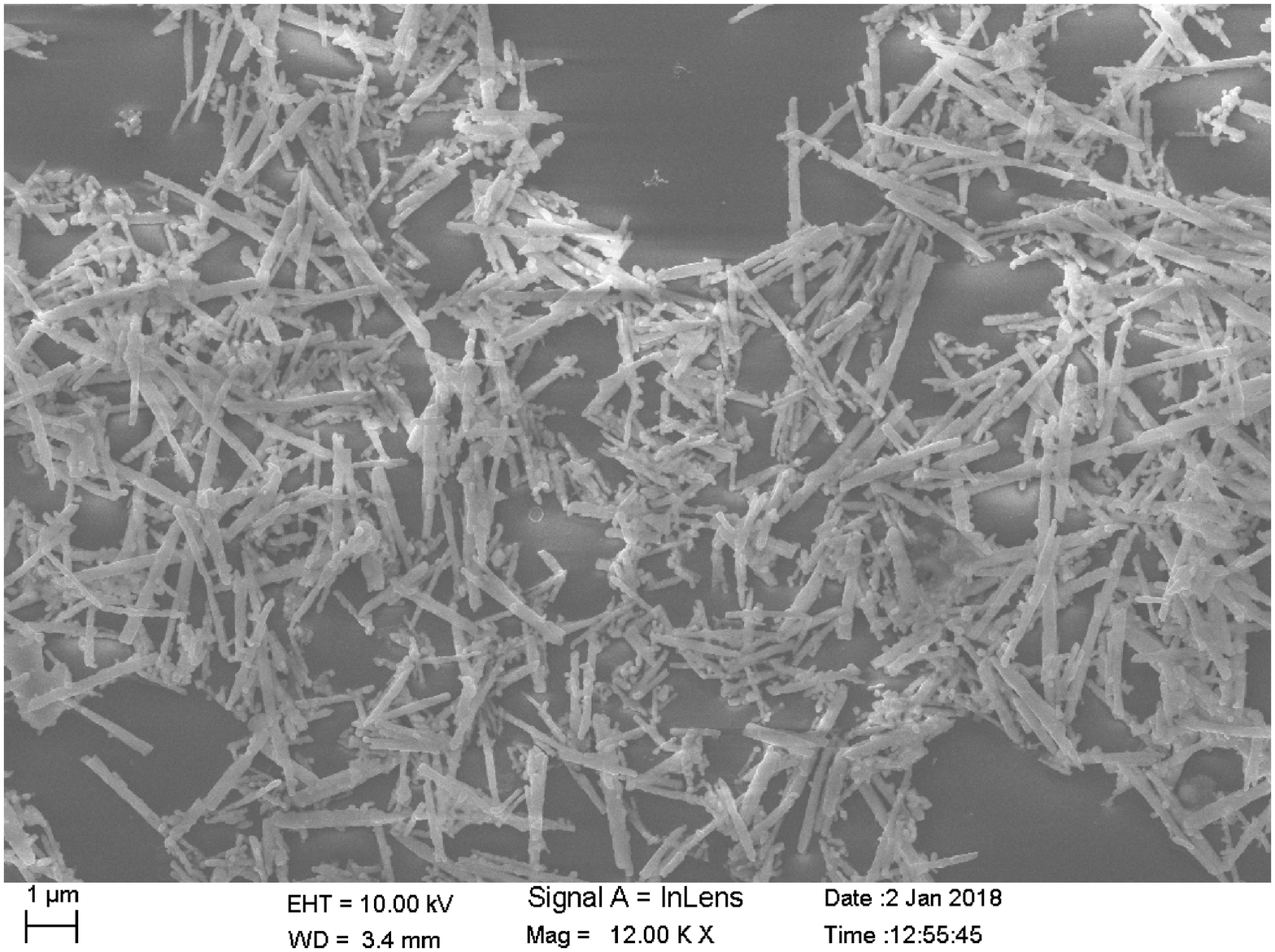}} 
   \subfigure[]{\includegraphics[scale=0.15]{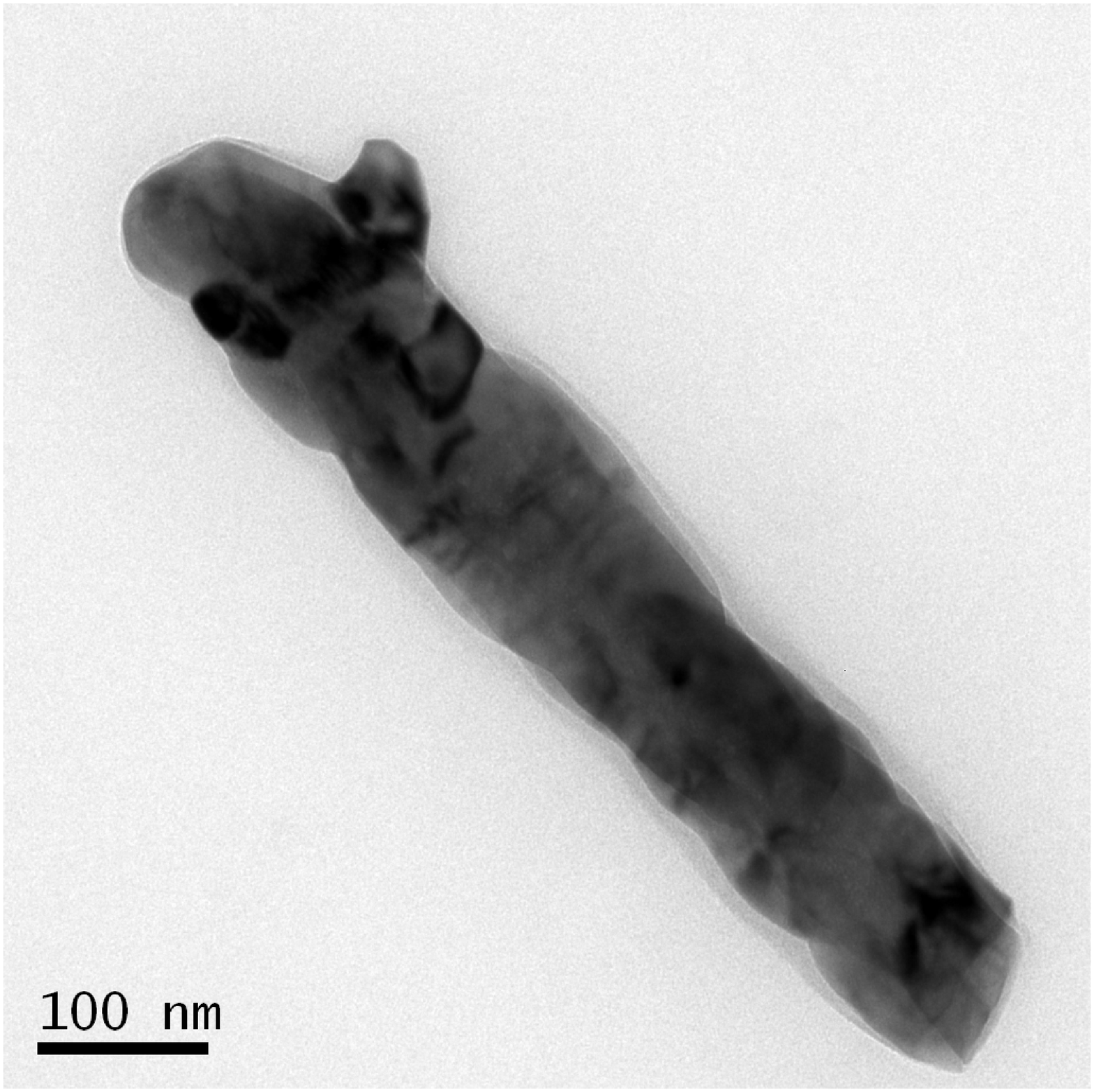}}
   \subfigure[]{\includegraphics[scale=0.15]{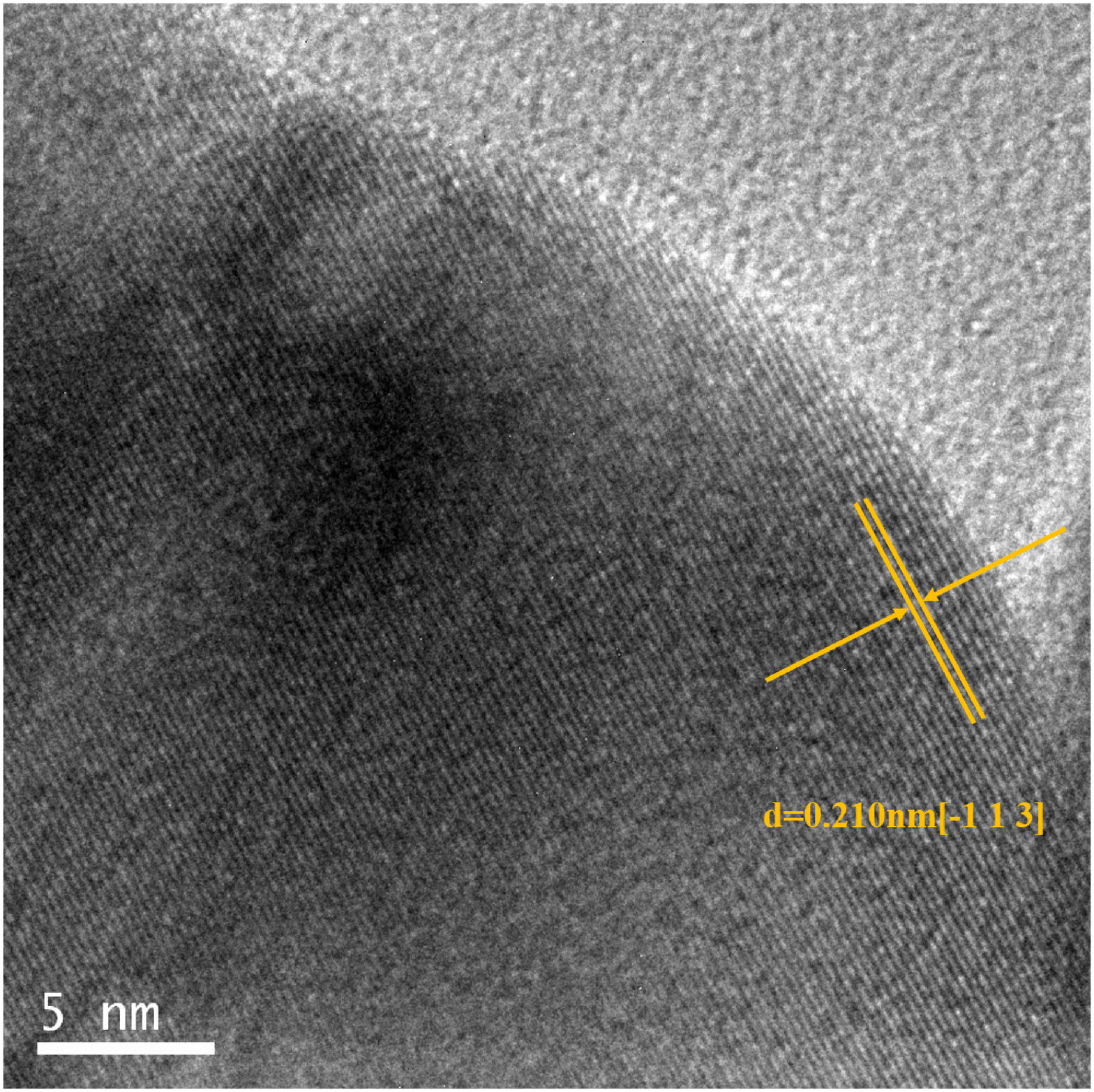}} 
   \subfigure[]{\includegraphics[scale=0.20]{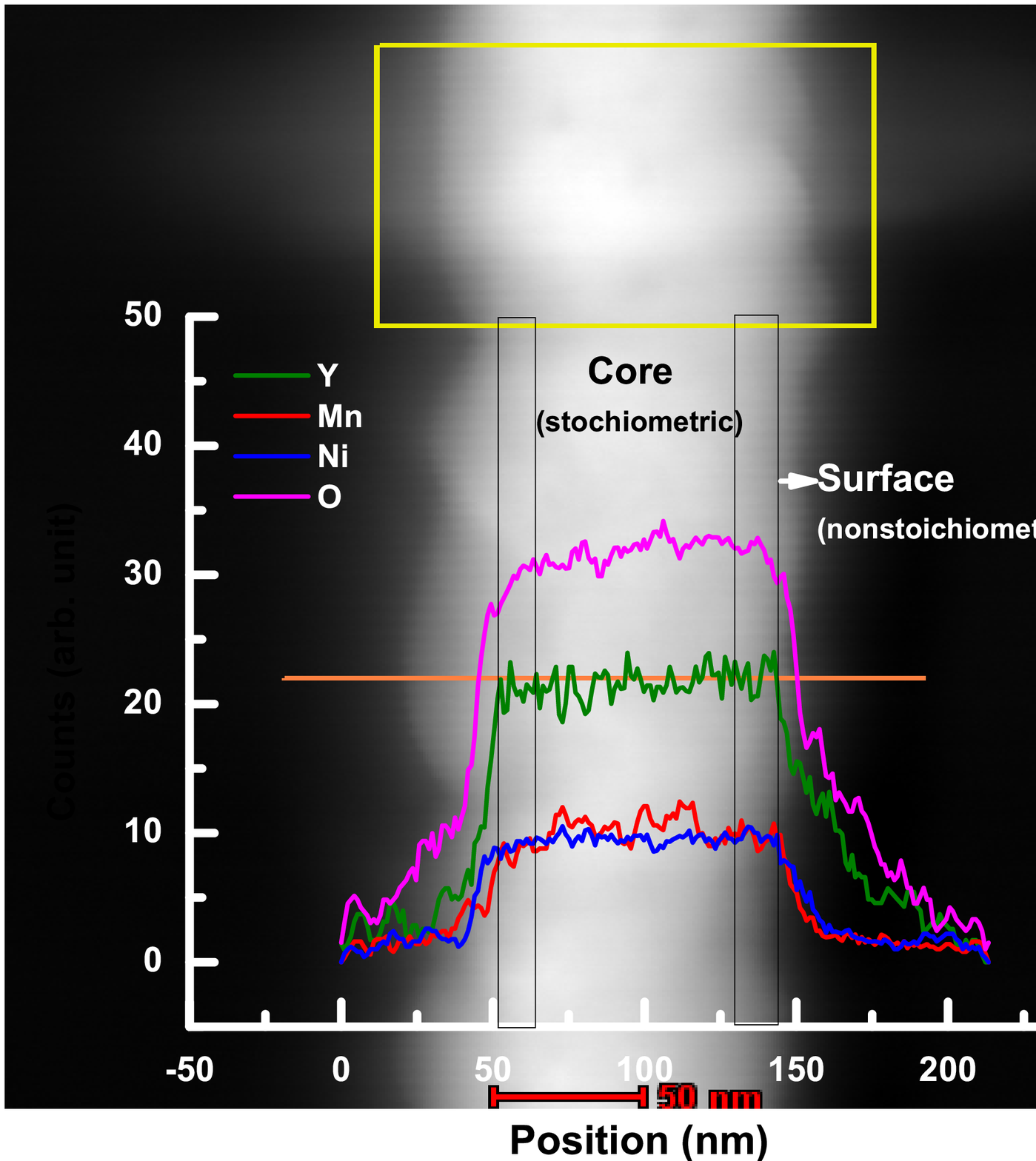}}
   \end{center}
\caption{The (a) field-effect scanning electron microscopy image of nanorods; (b) transmission electron microscopy image of a single nanorod; (c) high resolution TEM image showing the ($\bar{1}$13) planes; (d) HAADF image of the nanorod together with mapping of concentration of the elements such as Y, Mn, Ni, and O across the diameter of the nanorod as obtained from EDX line scanning; it helps in identifying the `stoichiometric' core and `nonstoichiometric' surface regions of the nanorod. }
\end{figure*}

\begin{figure*}[ht!]
\begin{center}
   \subfigure[]{\includegraphics[scale=0.25]{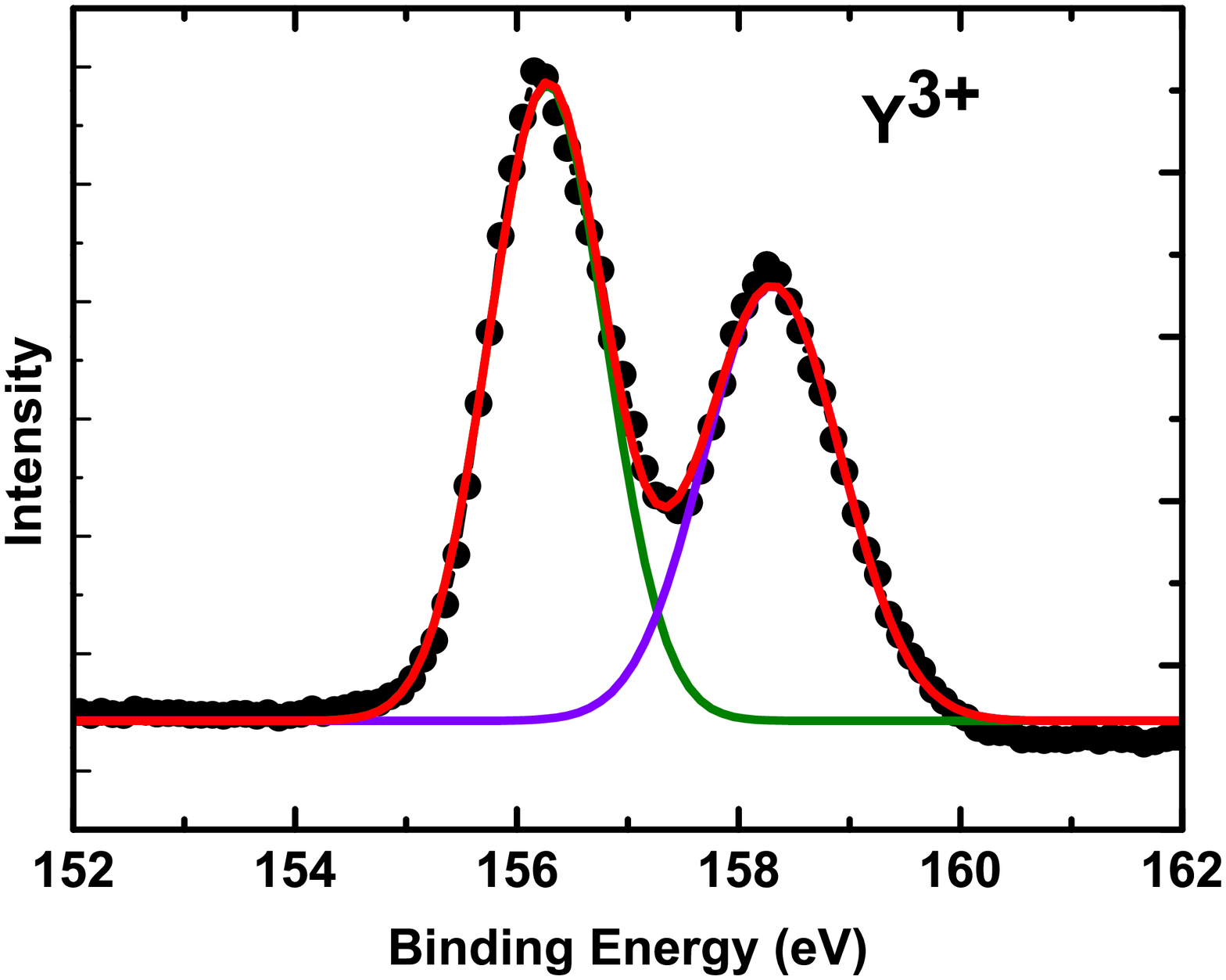}} 
   \subfigure[]{\includegraphics[scale=0.25]{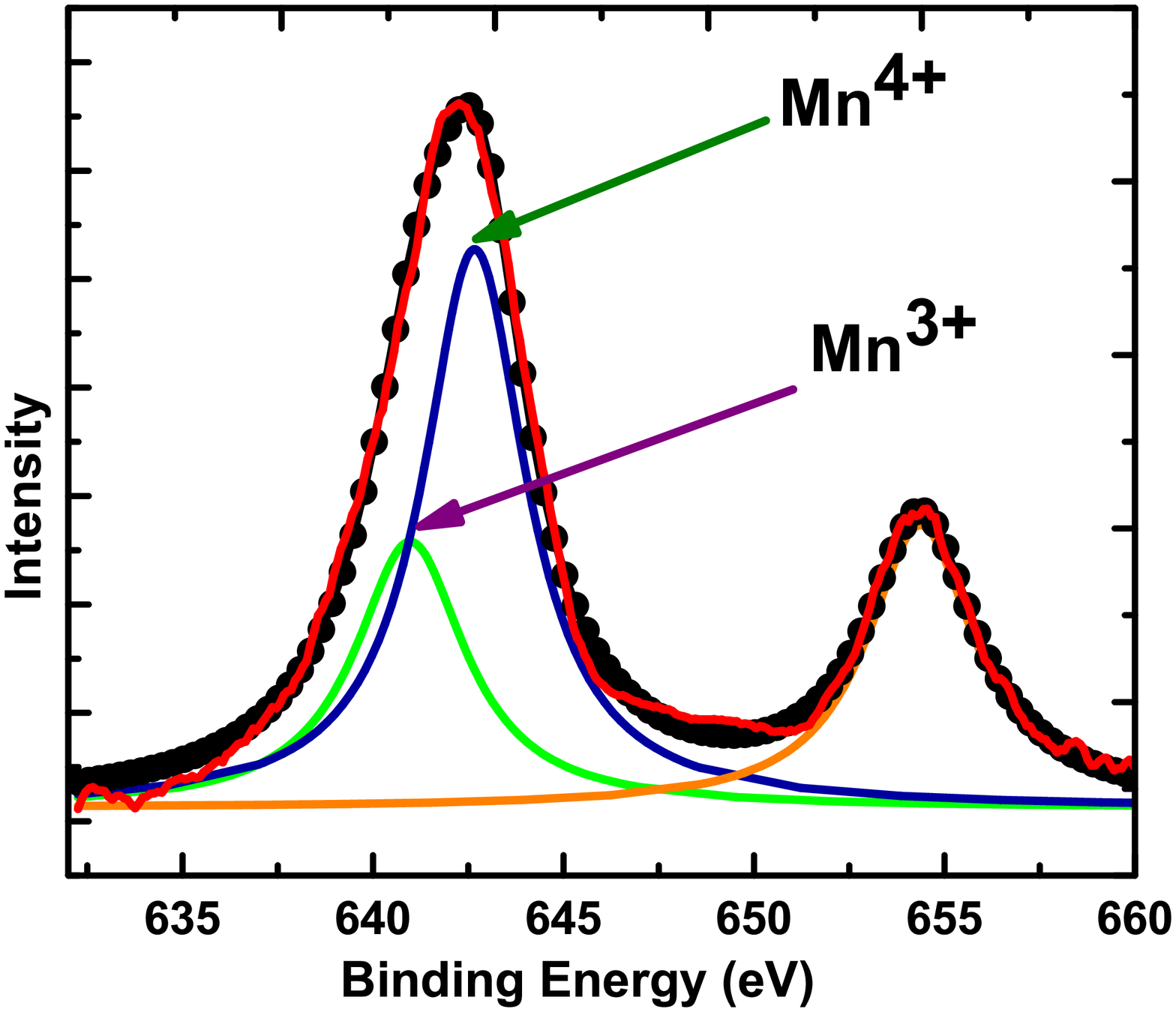}}
   \subfigure[]{\includegraphics[scale=0.25]{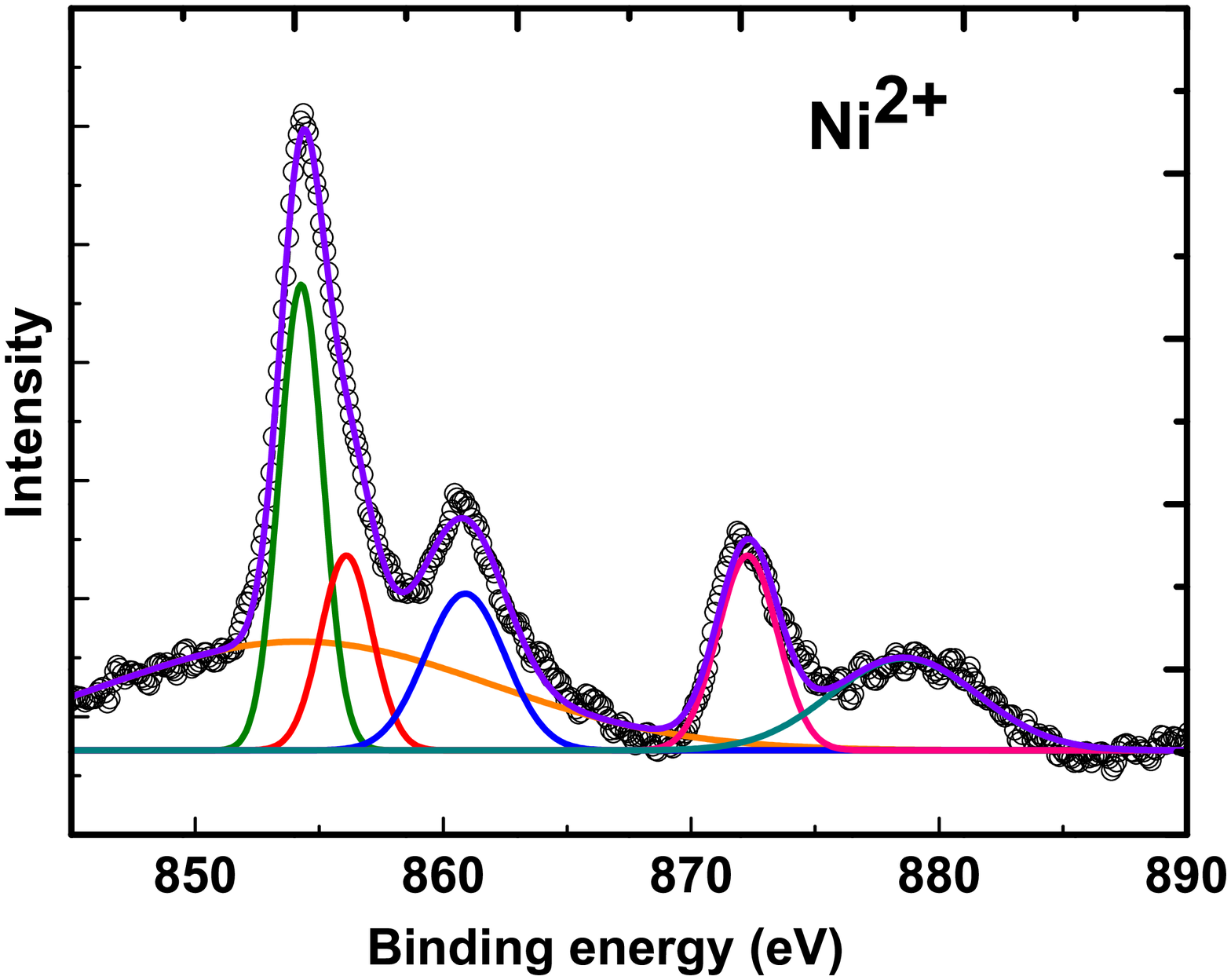}}
   \subfigure[]{\includegraphics[scale=0.25]{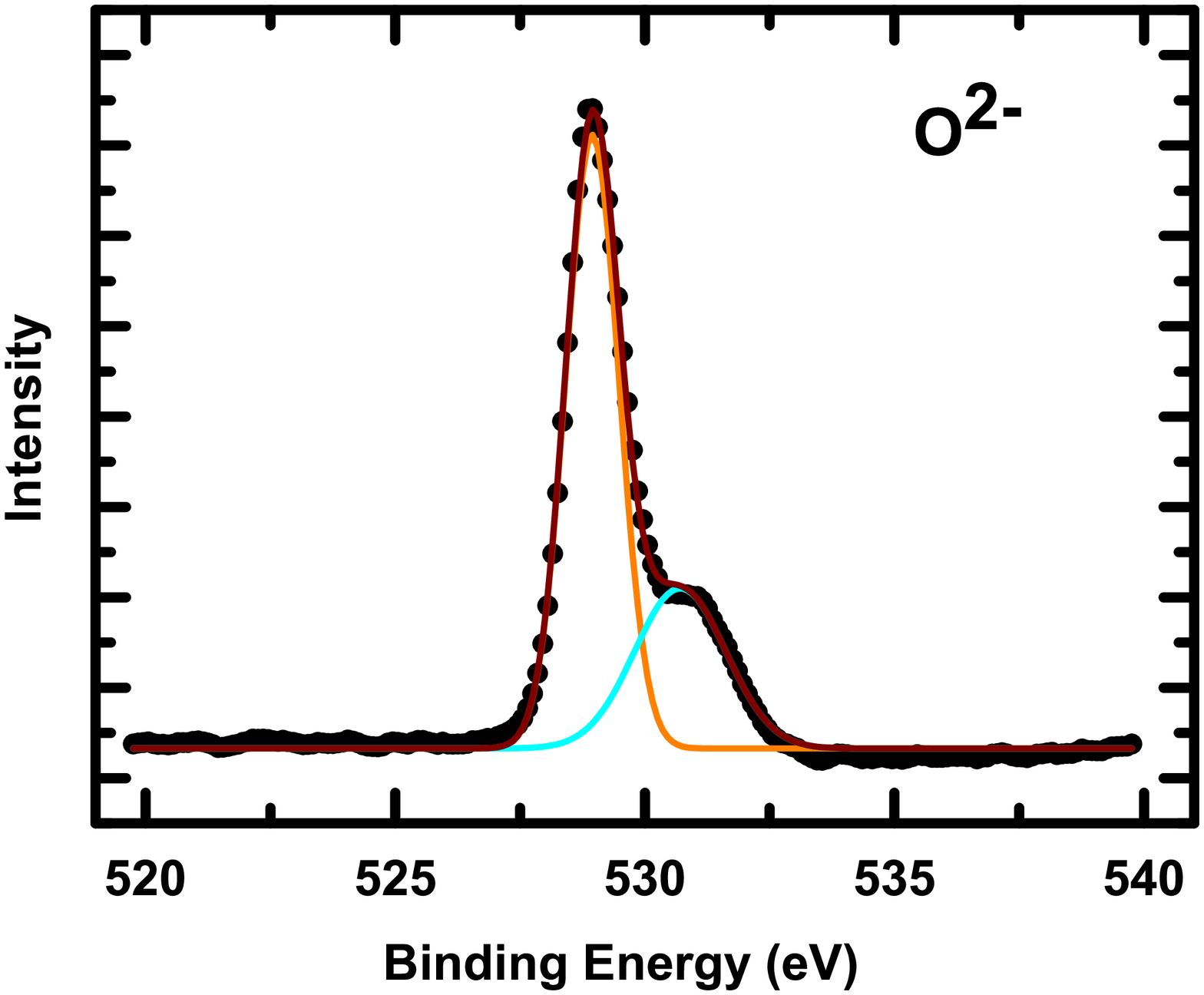}}
   \end{center}
 	\caption{The deconvoluted x-ray photoelectron spectra showing the core and satellite peaks for (a) Y, (b) Mn, (c) Ni, and (d) O and their fitting. }
\end{figure*} 

\begin{figure*}[ht!]
\begin{center}
   \subfigure[]{\includegraphics[scale=0.25]{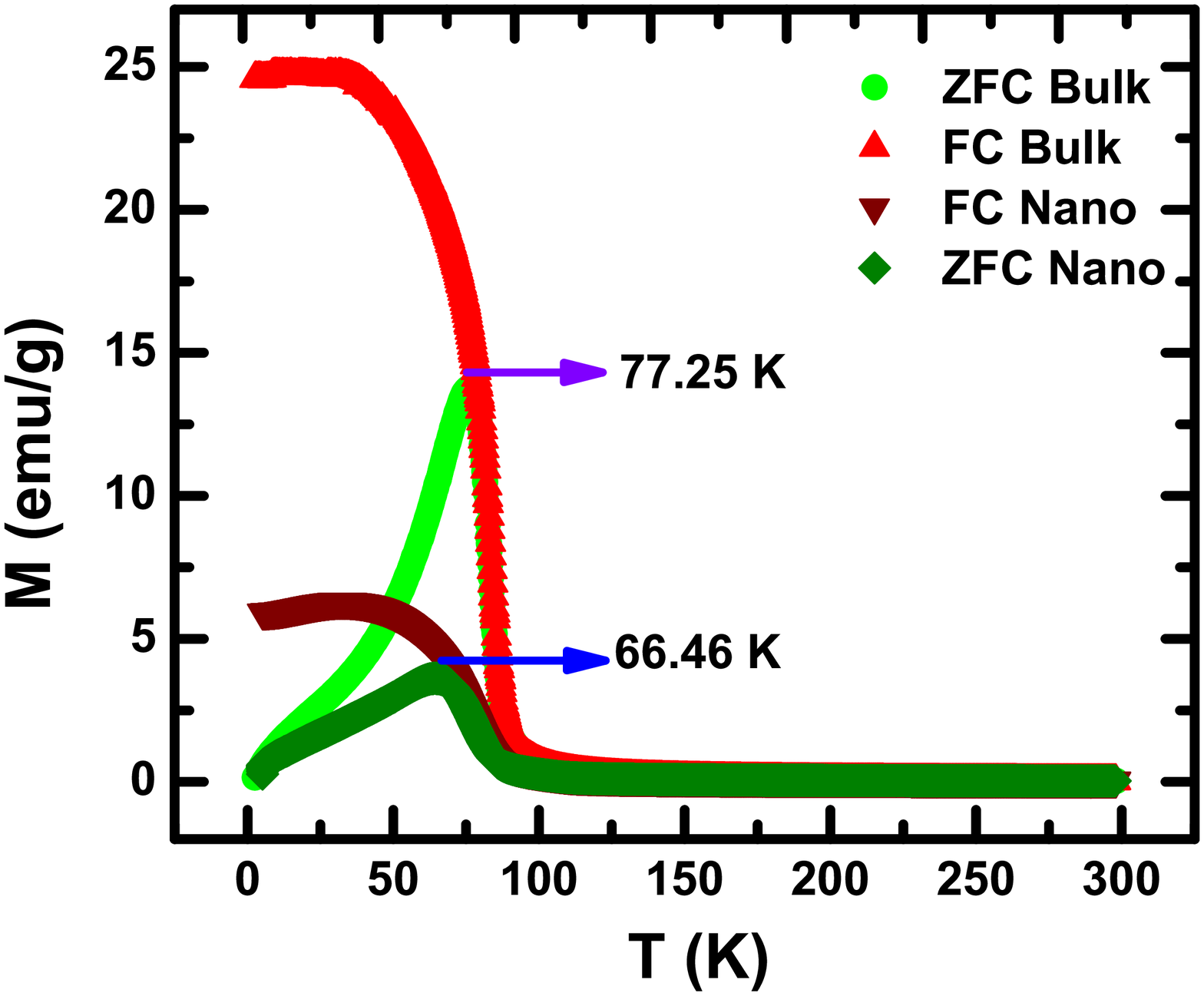}} 
   \subfigure[]{\includegraphics[scale=0.25]{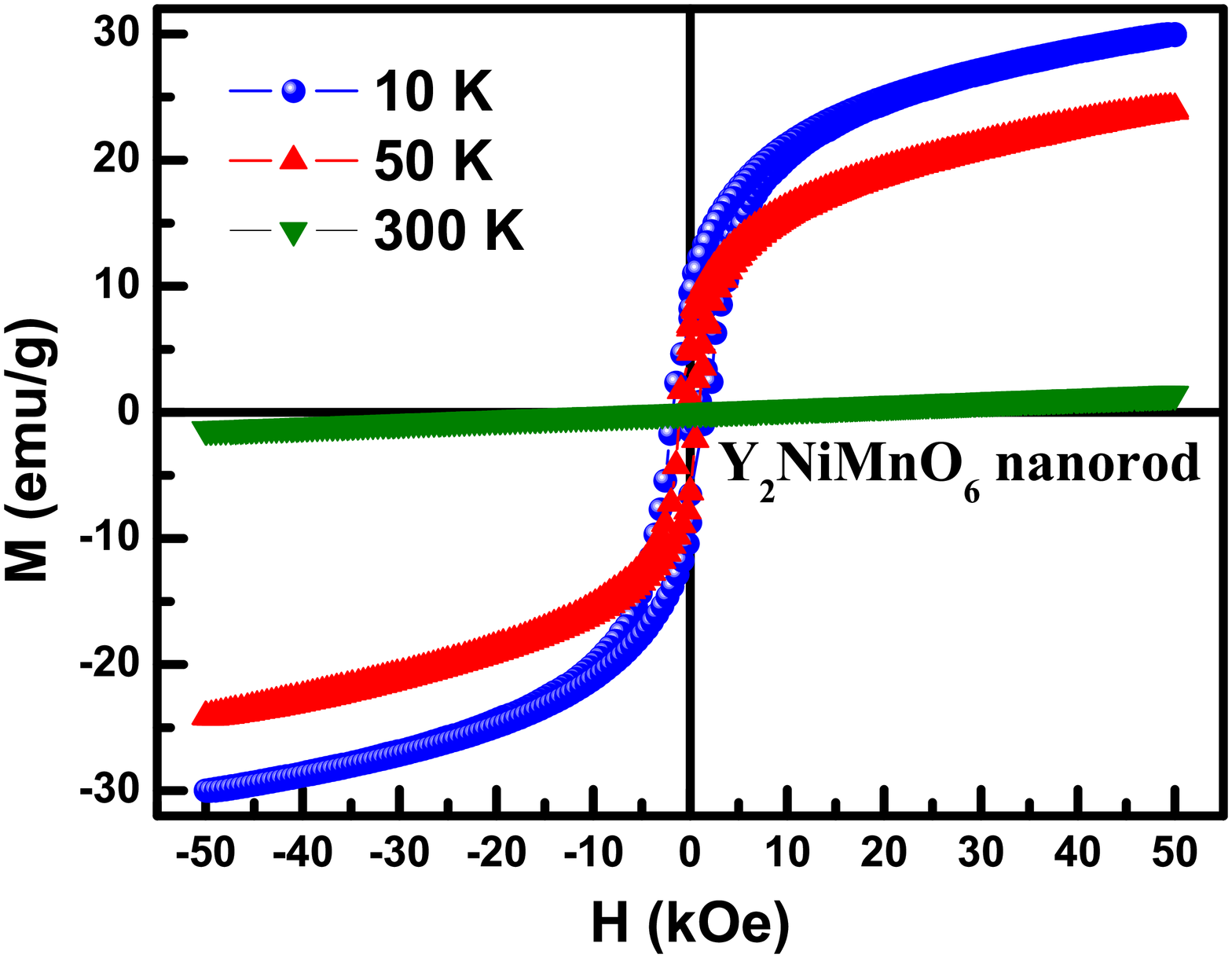}}
   \subfigure[]{\includegraphics[scale=0.25]{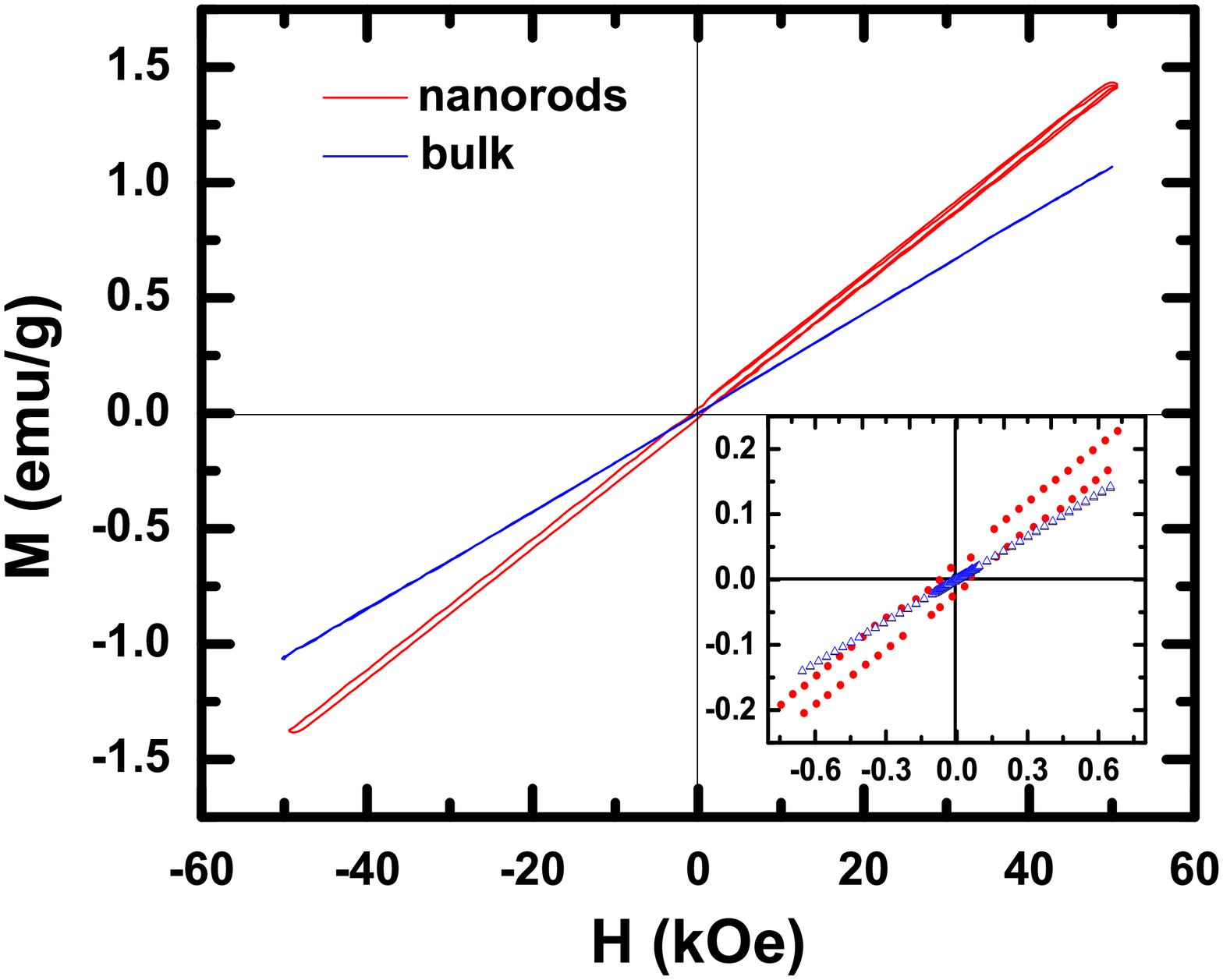}} 
   \subfigure[]{\includegraphics[scale=0.25]{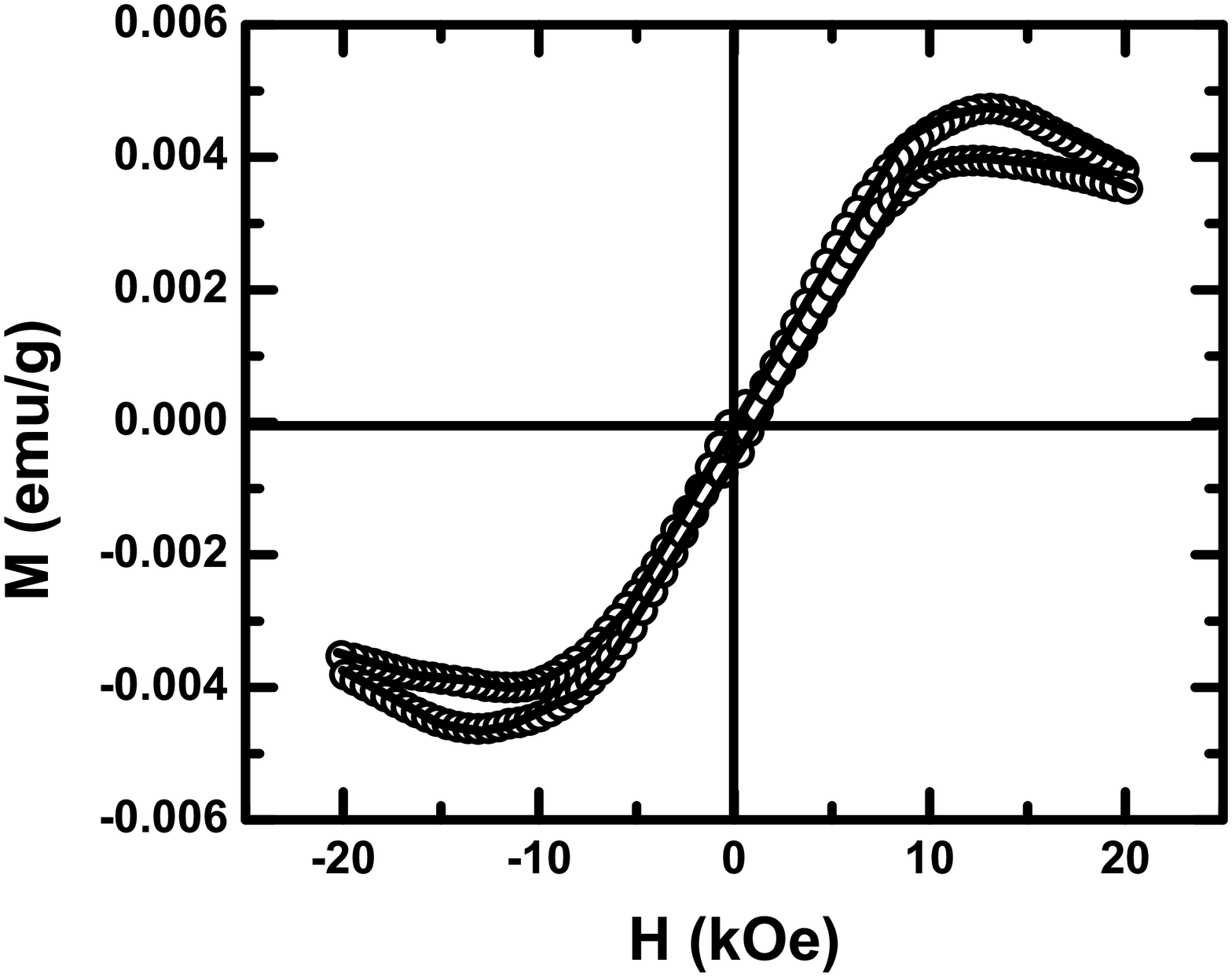}} 
   \end{center}
\caption{(a) The zero-field cooled and field-cooled magnetization versus temperature plots for bulk and nansocale YNMO; (b) magnetic hysteresis loops at low, intermediate, and room temperature for nanoscale YNMO; (c) magnetic hysteresis loops for bulk and nanoscale samples of YNMO at room temperature; inset shows the blown up portion of the loop near origin for both the samples; (d) weak nonlinear M-H pattern near origin, extracted from the overall M-H loop, signifying subtle surface ferromagnetism at room temperature in nanorods.}
\end{figure*}

\section{Results and Discussion}

The Figure 1 shows the room temperature x-ray diffraction patterns for bulk and nanorods of YNMO and their refinement by FullProf. The crystallographic structure is also shown in Fig. 1. The crystallographic structure is found to be monoclinic (space group $P2_1/n$) for both bulk and nanoscale samples though the nanorods appear to be oriented with ($\bar{1}$13) plane. It indicates a certain extent of texturing of the entire nanorod assembly on which the x-ray diffraction pattern was recorded. The lattice parameters were estimated to be $a$ = 5.221 \AA, $b$ = 5.553 \AA, $c$ = 7.479 \AA, $\beta$ = 89.87$^o$, and $a$ = 5.241 \AA, $b$ = 5.583 \AA, $c$ = 7.488 \AA, $\beta$ = 89.87$^o$, respectively, for bulk and nanoscale samples. Other crystallographic details such as ion positions, bond lengths and angles for both the bulk and nanoscale samples as well as the estimated standard deviation (corresponding to the lattice parameters and ion positions) and the fit statistics of the XRD data are included in the Supplemental Material \cite{Supplemental}. The preferential orientation of ($\bar{1}$13) plane could be observed in high resolution transmission electron microscopy (HRTEM) images as well (Fig. 2). The HRTEM images were analyzed by using fast fourier transformation (FFT) and its inverse (IFFT) in order to identify the lattice planes and vectors clearly. In the Supplemental Material \cite{Supplemental}, the HRTEM images and their FFT and IFFT versions are shown. The lattice  planes and vectors have also been shown. The growth axis [$\bar{1}$13] of the nanorods could be observed in the images (Supplementary Materials). The SEM and bright field TEM images are shown in Fig. 2. The diameter of the nanorods is found to be around 100 nm. We further carried out detailed energy dispersive X-ray (EDX) elemental line profile analysis across the diameter of an individual nanorod by using scanning TEM (STEM) high angle annular dark field (HAADF) technique inside TEM. The Figure 2(d) shows the HAADF image of the nanorod as well as a profile of elemental concentration of Y, Mn, Ni, and O across the diameter of the nanorod. Clearly, while all the other elements are found to be homogeneously distributed across the entire diameter of the nanorod, concentration of oxygen was found to be decreasing near and across the surface region. The profile of oxygen concentration across the diameter of the nanorod yields the thickness of the region to be $\sim$10 nm where oxygen vacancies form. We, therefore, divide the whole nanorod into two regions - `stoichiometric' and `nonstoichiometric' - as shown in Fig. 2(d). Using this information, it is possible to estimate the total number of unit cells within the `nonstoichiometric' region of a typical nanorod across which oxygen vacancies form. Considering the typical dimensions of the nanorod to be length $\approx$600 nm and diameter $\approx$100 nm and using the volume of the crystallographic cell ($\approx$0.21913 nm$^3$), the total number of unit cells in the whole of the nanorod is estimated to be 21504992. Since, the thickness of the `nonstoichiometric' region is $\approx$10 nm, the number of the unit cells within the `stoichiometric' region of the nanorod turns out to be 13763195. The number of cells within the `nonstoichiometric' region is $\sim$36\%. This is a siginificantly large number and, therefore, emergence of multiferroicity in this region should have profound impact. In order to corroborate the data of oxygen vacancy formation at the surface of the nanorods, we have also carried out the XPS. The spectra for Y, Mn, Ni, and O are shown in Fig. 3. The overall spectra were deconvoluted to obtain the main and sattelite peaks for each of the ions. For example, Fig. 3(a) shows the peaks at 156.5 eV and 158.5 eV for Y$^{3+}$ states. The Figure 3(b) shows the deconvoluted peaks - main and satellite - for Ni$^{2+}$. In both these cases, fitting of the spectra following background subtraction yields the charge states to be unique - i.e., Y$^{3+}$ and Ni$^{2+}$. Interestingly, however, Mn ions appear to be in the mixed state. Fitting of the peaks show that both Mn$^{3+}$ and Mn$^{4+}$ charge states are present. The ratio of the area ($A$) under the corresponding peaks [$A_{Mn^{3+}}$/($A_{Mn^{3+}}+A_{Mn^{4+}}$] is calculated to be $\sim$0.3. From the spectra for O$^{2-}$, the charge state of oxygen ion turns out to be -2. Because of oxygen deficiency in the surface region, the Mn$^{4+}$ ions were found to be reduced to Mn$^{3+}$ and thus, XPS data corroborate the observations made by TEM. It appears that $\sim$30\% of the Mn ions assume Mn$^{3+}$ state within a region consisting of $\sim$36\% cells. The XPS data for the bulk sample are shown in the Supplemental Material \cite{Supplemental}. All the peaks corresponding to Y$^{3+}$, Ni$^{2+}$, Mn$^{4+}$, O$^{2-}$ ions are found to reflect unique charge states. Signature of the presence of mixed charged state could not be observed in this case. This observation is consistent with those made by others for bulk YNMO \cite{Su}. Only in nanoscale samples, Mn$^{3+}$ arises because of oxygen vacancies near the surface region.

We now turn our attention to the results of magnetic property measurements. The zero-field-cooled (ZFC) and field-cooled (FC) magnetization versus temperature data are shown in Fig. 4(a) for both the bulk and nanoscale YNMO. While bulk sample exhibits the antiferromagnetic transition at $T_N$ $\approx$ 77 K, as expected, in the nanoscale samples, the $T_N$ is found to have dropped to $\sim$66 K. This is consistent with the observations made in other nanoscale magnetic compounds \cite{Zhao}. Interestingly, within this size range, the nanorods appear to retain the long-range magnetic order (albeit with weaker strength) in the bulk of the sample. The Figures 4(b) and (c) show, respectively, the magnetic hysteresis loops across $\pm$50 kOe for nanoscale YMNO at different temperatures across 10-300 K and those for both bulk and nanoscale samples at 300 K. Clearly, the bulk sample exhibits linear i.e., paramagnetic $M-H$ pattern at 300 K. In contrast, the nanoscale sample exhibits finite magnetic coercivity [Fig. 4(c) inset] and very weak nonlinearity because of subtle surface magnetism. Using Brillouin function for the nonlinear ferromagnetic component and linear function for the paramagnetic one, we extracted the ferromagnetic component of the overall magnetization observed experimentally [Fig. 4(d)]. It yields the saturation magnetization $M_S$ to be $\sim$0.005 emu/g at room temperature. This turns out to be $\sim$5\% of the maximum ferromagnetic moment expected for perfect `ferro' alignment of all the spins in YNMO. Near linearity in $M-H$ also signifies presence of long-range antiferromagnetic order in the surface region. The coercive fields for forward and reverse branches turn out to be $H_{C1}$ $\sim$735 Oe and $H_{C2}$ $\sim$600 Oe, respectively. This yields the cocercivity $H_C$ $\sim$666 Oe and exchange bias field $H_E$ $\sim$67.5 Oe. Finite $H_E$ confirms presence of surface antiferromagnetism and interface between ferromagnetic and antiferromagnetic regions. The shape of the hysteresis loop, though appears to be unusual, resembles closely the one expected of a dilute assembly of noninteracting three-dimensional nanoparticles with higher packing density \cite{Usov}.

\begin{figure*}[ht!]
\begin{center}
   \subfigure[]{\includegraphics[scale=0.30]{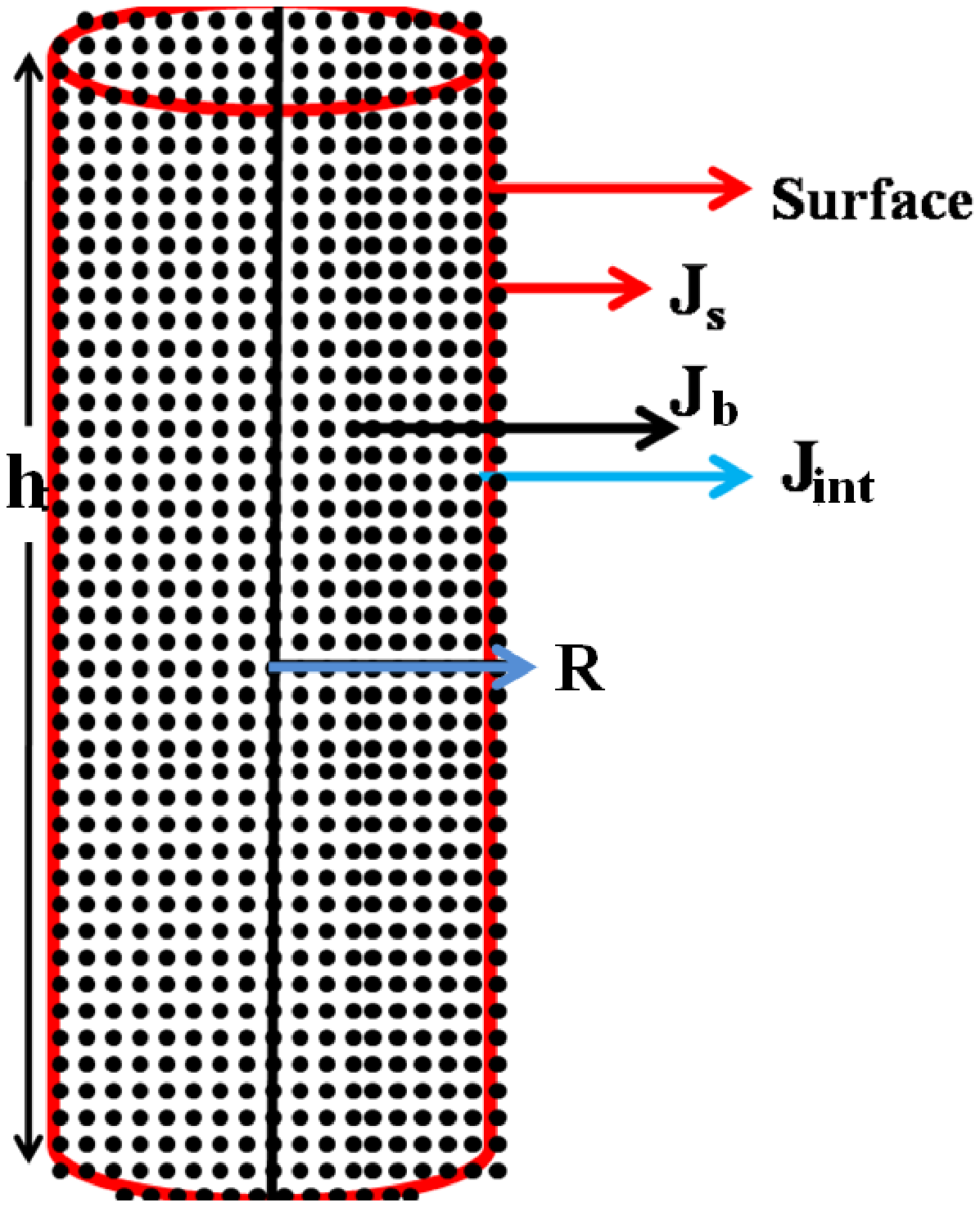}} 
   \subfigure[]{\includegraphics[scale=0.15]{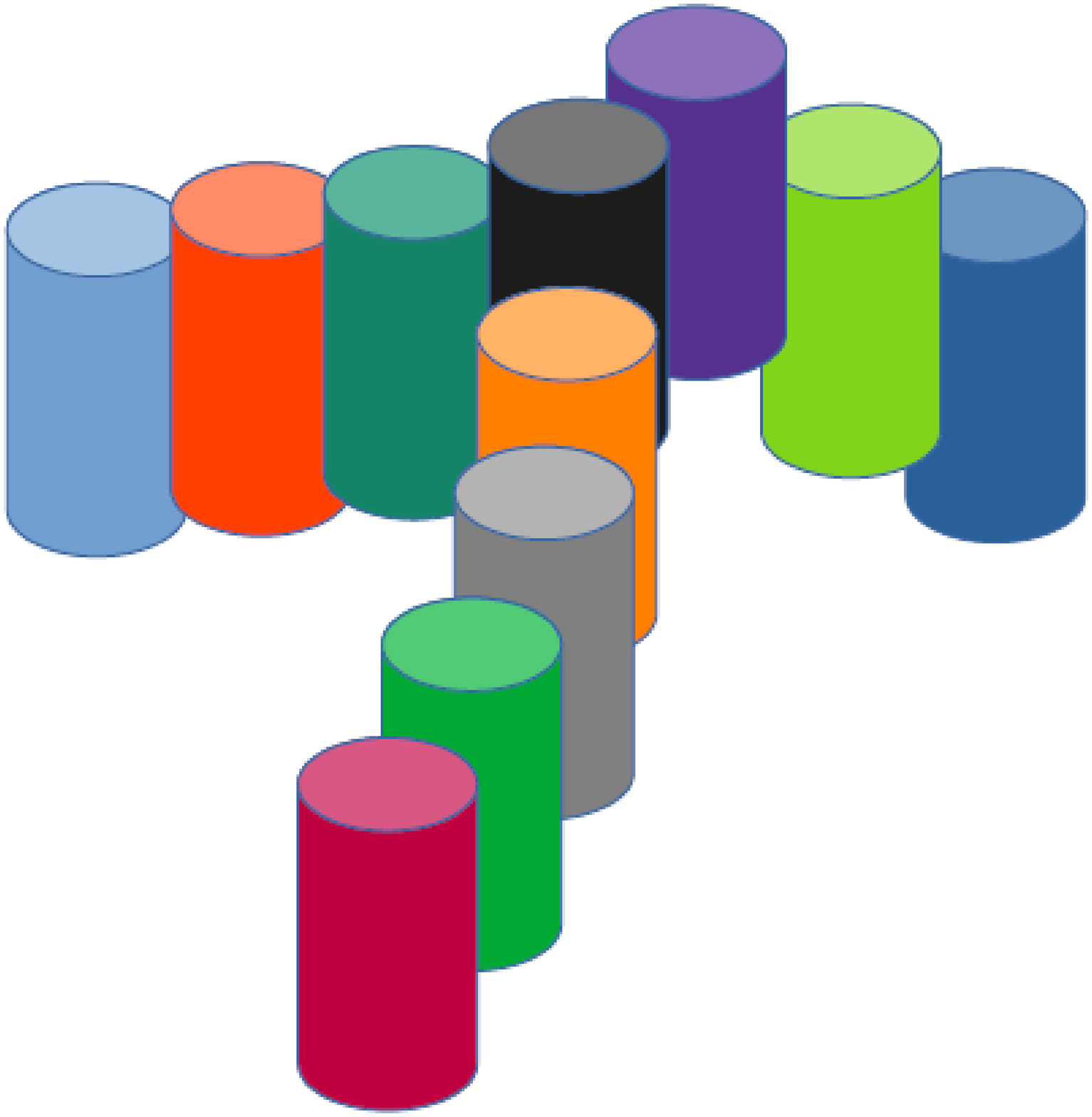}}
   \subfigure[]{\includegraphics[scale=0.32]{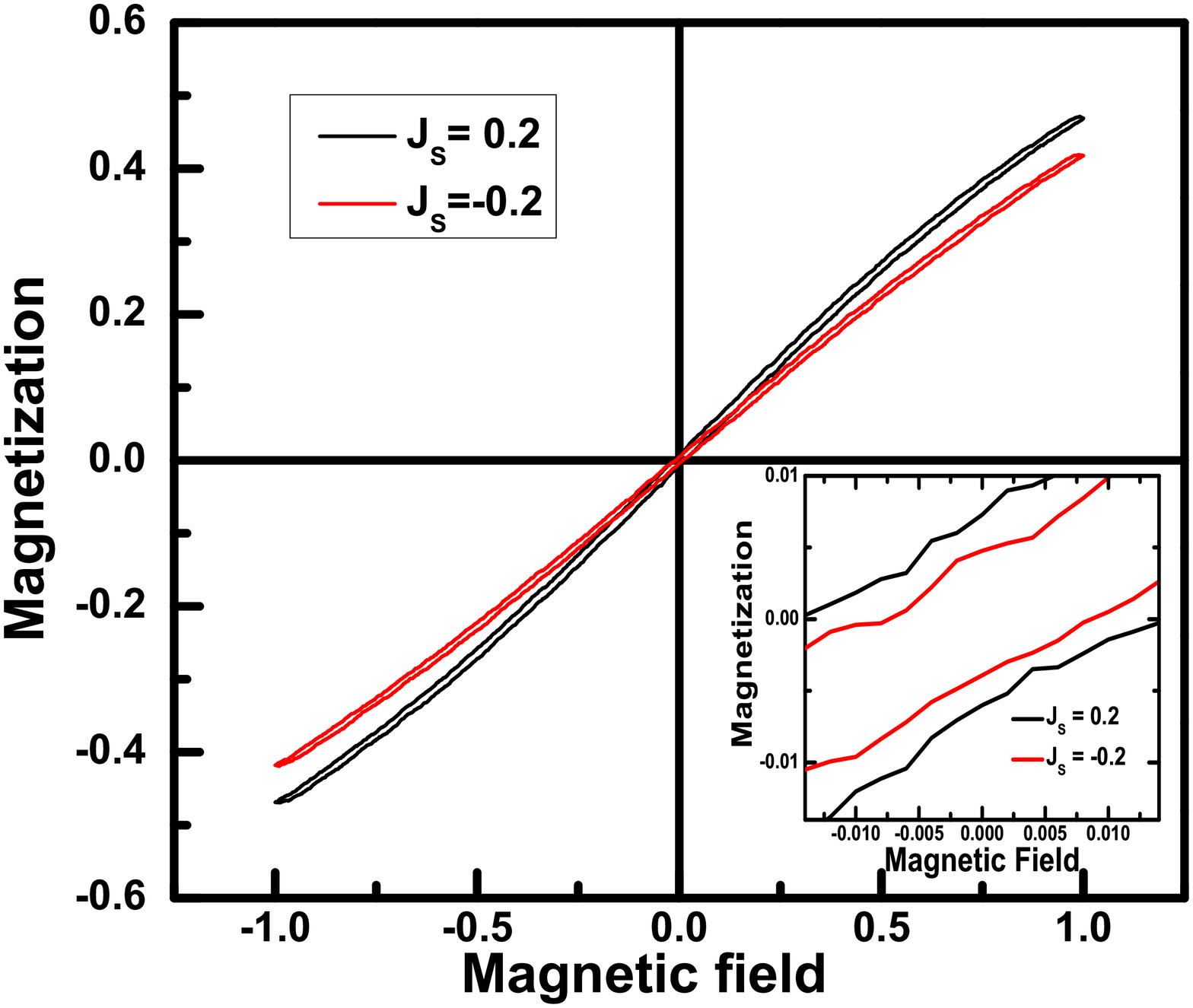}}  
   \end{center}
\caption{(a) Schematic of the cross-sectional view of nanorod of radius radius $R$ and height $h$. The Heisenberg spins interact at the surface (inner core) of the rod by the strength $J_s$ ($J_b$) and the interaction strength across the interface is $J_{int}$; (b) a randomly assembled structure; (c) M-H loop with $J_s = 0.2$ and $J_s = -0.2$.}
\end{figure*}

To examine the origin of magnetic coercivity of the assembled nanorods we 
consider that the surface of the nanorods are ferromagnetic in nature. We model the rod-like  nanoparticles as cylinders  of radius $R$ and height $h$ [as shown in Fig. 5(a)]  consisting of classical Heisenberg spins ${\bf S}_i$ arranged in a cubic lattice \cite{Sahoo}, i.e., there  are $2R + 1$ spins along the diameter of the cylinder; $N$ such identical nanorods are assembled randomly to form the superstructure as shown in Fig. 5(b) which   mimics the self-assembled nanorods observed in our experiments. For comparison, we also consider two other kind of assmbled structures by attaching the rods as {\it end to end} and {\it side by side}, which is shown in the Supplemental Material \cite{Supplemental}.

Each lattice site $i$ of the superstructure is associated with a classical Heisenberg spin ${\bf S}_i$; we denote the lattice sites in the bulk by a set {\bf B} and  those on  the surface  by a set {\bf S}. These  spins interact following the Hamiltonian, 
\bea
 {\cal H} &=&   -J_b \sum_{i\in {\bf B}, j\in {\bf B}} {\bf S}_i. {\bf S}_j  -J_{s} \sum_{i\in {\bf S}, j \in {\bf S}}  {\bf S}_i. {\bf S}_j \cr &&-  J_{int}\sum_{i \in {\bf B},  j \in {\bf B}} {\bf S}_i. {\bf S}_j   - H \sum_{i\in {\bf B}, i\in {\bf S} } S_i^z,
 \label{eq:H}
 \eea
where $j$ is the nearest neighbor of site $i$,  $J_{b}$, ($J_{s}$) are the exchange interaction strength among the spins within the bulk {\bf B} (surface {\bf S}); $J_{int}$  represents interaction between spins in the bulk and  the surface. The external magnetic field $H$ is applied along the easy-axis, which is chosen as the $z$-direction. To model a paramagnetic core (bulk) and ferro or antiferromagnetic surface we set  $J_{b}$ to be  very small  and $J_{s}$ can take  positive or negative values depending on whether the surface is ferro or antiferromagnetic, respectively.

We study the hysteresis properties of these nanoparticle super-structures using Monte Carlo simulations with a single spin-flip Metropolis algorithm where a trial configuration is accepted  with probability  $ Min\{ 1, e^{-\beta \Delta E}\}$. Here, $\Delta E$ is energy difference between the present configuration and the trial one. The trial configuration is constructed by changing the angles of a randomly chosen spin by a small but random amount. 

To calculate the hysteresis  of the self-assembled nanorods we consider $N=11$ rods arranged randomly, as shown in Fig. 5(b). The radius $R$  and the height  $h$ of the nanorods are taken as $8$  and $30$ lattice units respectively. The   interaction parameters of the Hamiltonian are set as follows. For the paramagnetic core and ferro or antiferromagnetic surface, we set  $J_{b}=0.01$  and  consider   $J_{s}$  in the range $(-0.8, 0.8)$; $J_{int}$ is  expected to be similar in magnitude as that of $J_b$ and we set $J_{int}=0.01$. The temperature of the system is set as  $\beta^{-1}=1$, which is much smaller than the critical temperature $T_C$ of the system; such a state with $\beta^{-1}= 1$ can be achieved in the Monte Carlo simulation, starting from any random initial configuration, by relaxing the system in zero-field-cooled condition for a long time. Then, the magnetic field  is raised slowly from $H=0$ to $H_{max}$ with a field sweep rate $\Delta H$ units per Monte Carlo sweep (MCS) and finally, the hysteresis loop calculations are undertaken for a cycle by varying the field from $H_{max}$ to $ -H_{max}$ and then to $ H_{max}$. To this end we take  $H_{max}=1$ and $\Delta H$ = $2\times10^{-3}$, i.e., the magnetic field is raised from $0$ to $1$ in 500 MCS.   

We have calculated the hysteresis  loop for  different  $-0.8<J_s <0.8$; the magnetization loop for  $J_{s} = 0.2$, shown in Fig. 5(c), has {\it linear} behaviour and a small coercive field, which  compares well with our experimental data. This perhaps confirm the presence of ferromagnetic ordering at the surface of the self-assembled  nanorods along  with a paramagnetic ordering  in the bulk. For comparison, we, in Fig. 5(c), have also shown  the hysteresis loop for  $J_{s} = -0.2$, where  the surface has a antiferromagnetic order. Higher coercivity [Fig. 5(c)] for $J_s$ = 0.2 corroborates the experimental result. It proves presence of surface ferromagnetism in the present case.

\begin{figure*}[ht!]
\begin{center}
   \subfigure[]{\includegraphics[scale=0.22]{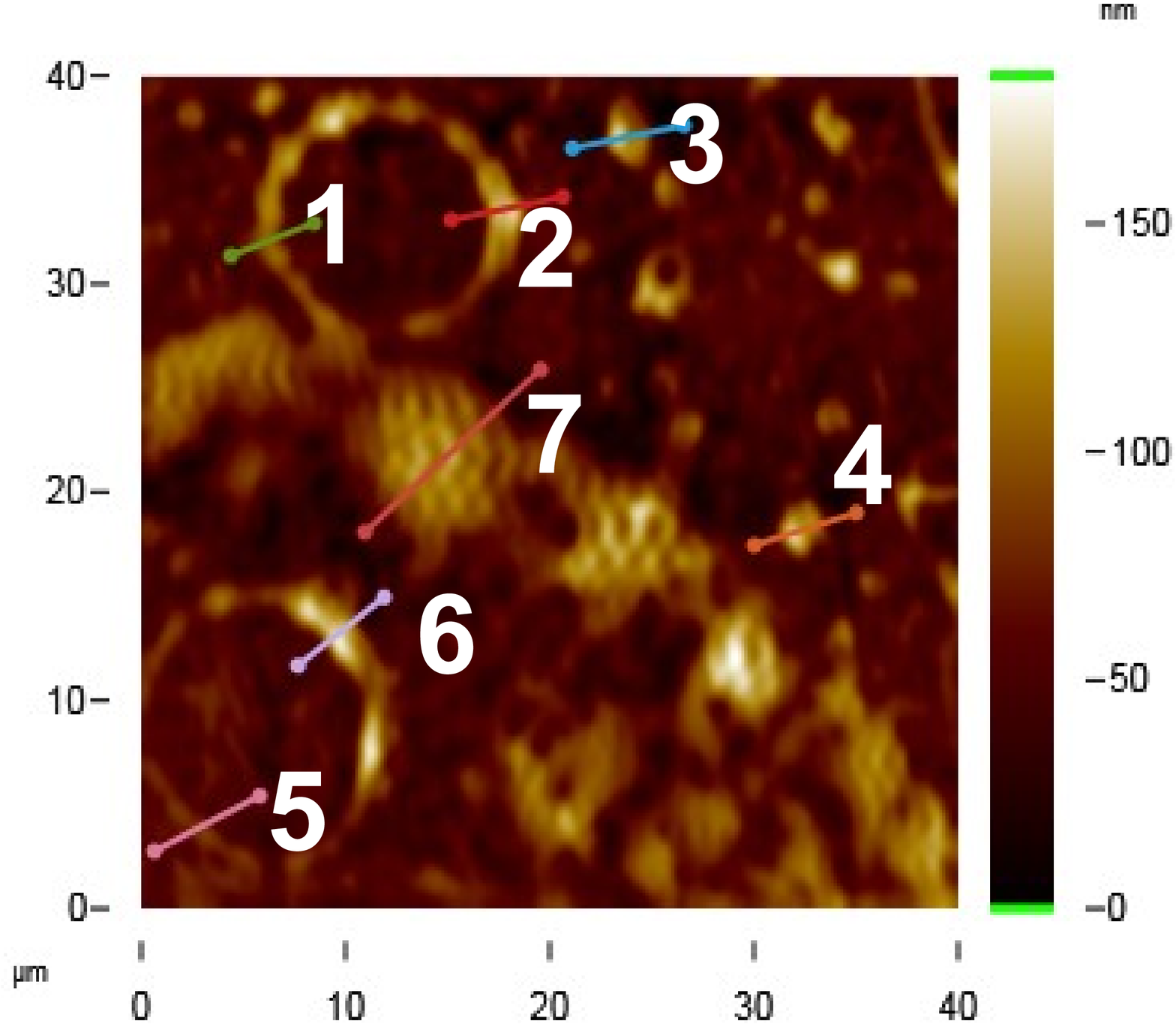}} 
   \subfigure[]{\includegraphics[scale=0.22]{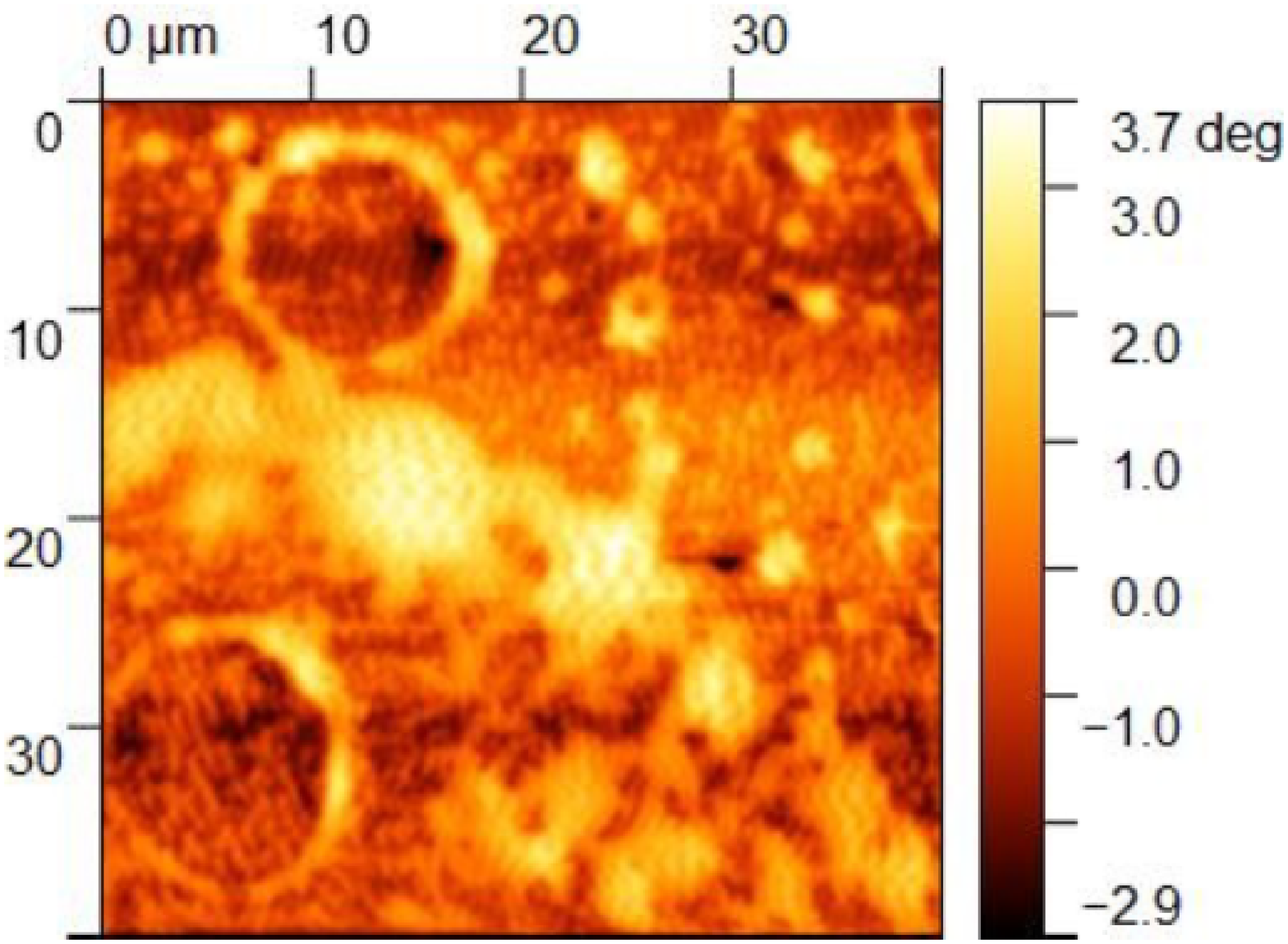}}   
   \subfigure[]{\includegraphics[scale=0.22]{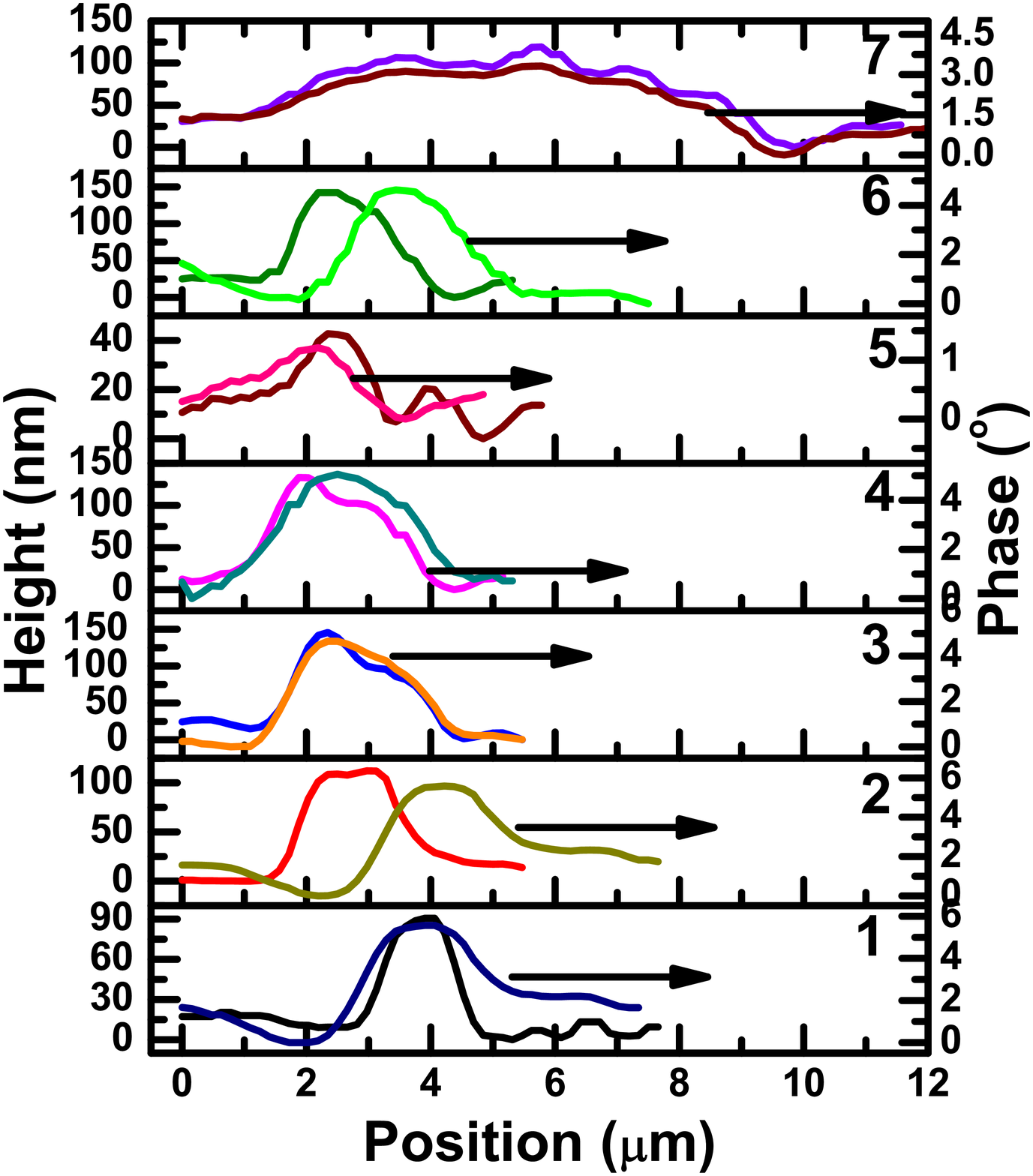}}
   \subfigure[]{\includegraphics[scale=0.22]{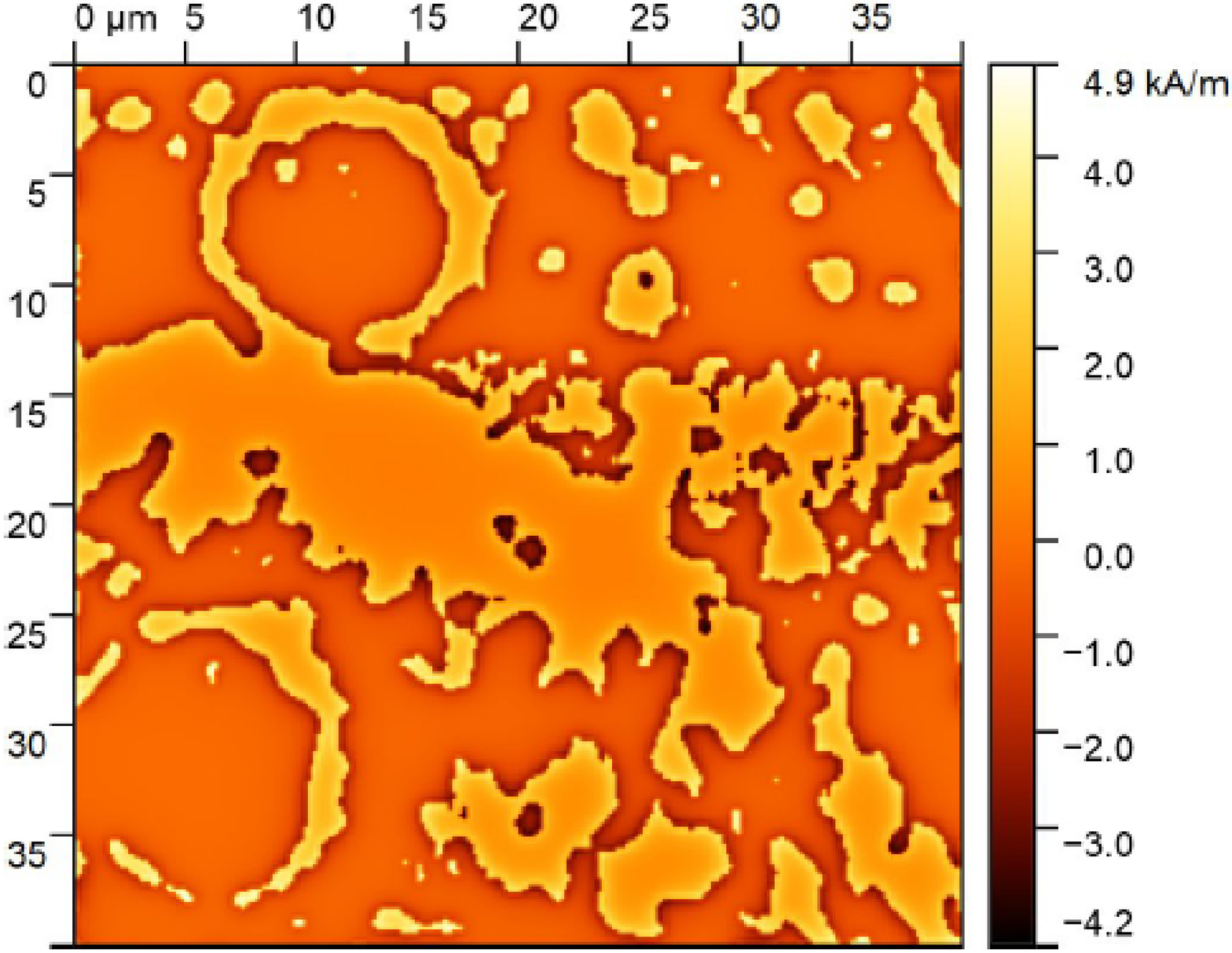}}
   \end{center}
\caption{The magnetic force microscopy images show the (a) surface topography and (b) phase contrast; (c) corresponding line profile scan data for the cluster of nanorods are shown across the lines shown in the images; (d) shows the mapping of the perpendicular stray field; line profile scan data corresponding to the phase contrast image have been generated from this analysis.}
\end{figure*}

The room temperature surface ferromagnetism is probed further by magnetic force microscopy (MFM). We show, respectively, in Figs. 6(a) and 6(b) the topography and MFM phase contrast images of nanorods of YNMO dispersed within ethanol and deposited on SiO$_2$/Si substrate. The sample was kept under $\sim$1000 Oe field prior to the measurements. The particles, as a result, appear to organize themselves in the form of rings via, possibly, magnetic interactions. The line profile data corresponding to the topography and phase contrast images are shown in Fig. 6(c). The phase contrast image [Fig. 6(b)] is also analyzed by using Gwyddion software for mapping the distribution of stray field. The details of the analysis is described in the Supplemental Material \cite{Supplemental} (see also the reference [S1] therein). The distribution of the stray field is shown in Fig. 6(d). The line profile data corresponding to the phase contrast MFM image have, in fact, been generated from this analysis. The comparison of the line scan data for topography and phase contrast images shows how the perpendicular stray field extends across the cluster of nanorods in different zones of the area under focus. From this, it appears that the magnetic domains extend across the individual nanorods and correspond closely to the respective cluster size as observed in many such magnetic nanoparticle assemblies \cite{Puntes}. All these information confirm the presence of weak yet finite surface ferromagnetism as paramagnetic strucure would not have resulted in such stray field distribution across the particle surface. The observation of long-range magnetic order and oxygen deficiency within the surface region, therefore, points out that Ni$^{2+}(3d^8; S=1)$-O$^{2-}(2p^6; S=0)$-Mn$^{3+}(3d^4; S=2)$-O$^{2-}(2p^6; S=0)$-Mn$^{4+}(3d^3; S=3/2)$ pathway is responsible for developing the exchange coupling interactions in the surface. We next examine the exchange coupling interaction in detail. The surface ferromagnetism, in many oxide perovskite systems, was shown \cite{Aliyu} to result from double exchange interactions involving mobile charge carriers arising from oxygen deficiency at the surface. Identification of the appropriate exchange interaction pathway is important. For single transition metal ion systems, it is straightforward. However, for systems involving more than one transition metal ion, as in our case, the exchange interaction pathway could be complex and, therefore, the mechanism too could be more involved. In the bulk YNMO, the exchange coupling interactions across Ni$^{2+}$-O$^{2-}$-Ni$^{2+}$, Mn$^{4+}$-O$^{2-}$-Mn$^{4+}$, and Ni$^{2+}$-O$^{2-}$-Mn$^{4+}$ bonds were found to be responsible to yield E-type antiferromagnetic structure $\uparrow\uparrow\downarrow\downarrow$ below $\sim$70 K. In the case of nanorods, because of the presence of Mn$^{3+}$ ions at the surface, exchange coupling interactions across Ni$^{2+}$-O$^{2-}$-Mn$^{3+}$, Mn$^{3+}$-O$^{2-}$-Mn$^{3+}$, Mn$^{3+}$-O$^{2-}$-Mn$^{4+}$ bonds yield the surface magnetism at room temperature. Within the surface crystallographic structure (which because of its lower dimension and consequent noncentrosymmetry could yield large Rashba spin-orbit coupling), the DM exchange across Mn$^{3+}$-O$^{2-}$-Mn$^{3+}$ could stabilize. 

It is, of course, important to mention here that the symmetry breaking magnetic structure could emerge either due to Dzyaloshinskii-Moriya (DM) antisymmetric exchange coupling interaction among the noncollinear spins or from exchange striction interaction among the collinear spins \cite{Nagaosa}. The spin-orbit coupling assumes immense importance in the former case but not in the latter. The theoretical and experimental work, carried out so far \cite{Fert}, have, of course, shown quite convincingly that the lower dimensional structures - such as surface and interface regions - or artificially constructed two-dimensional layers exhibit stabilization of noncollinear spin structures with DM exchange interaction. The role of Rashba spin-orbit coupling has been examined in such lower-dimensional structure in detail \cite{Fert}. Symmetric ``exchange striction" interaction may not be quite relevant in such systems. Direct experimental proof \cite{Gross,Chaurasiya} of prevalence of DM exchange interaction in the cases of surface/interface magnetism has also been provided. Since, in the present case, we observe ferroelectricity originating from surface magnetism at room temperature (where bulk magnetism has no role to play), it is quite likely that the surface magnetism in this case of Y$_2$NiMnO$_6$ nanorods involves noncollinear spin structure. Therefore, ferroelectricity here should be originating from the antisymmetric DM exchange coupling interaction (involving large Rashba spin-orbit coupling) confined within the lower dimensional surface region of the nanorods identified by the TEM experiments. The exchange striction interaction may not play any significant role here. This issue will be taken up further by using first-principles calculations separately.

\begin{figure}[ht!]
\begin{center}
   \subfigure[]{\includegraphics[scale=0.25]{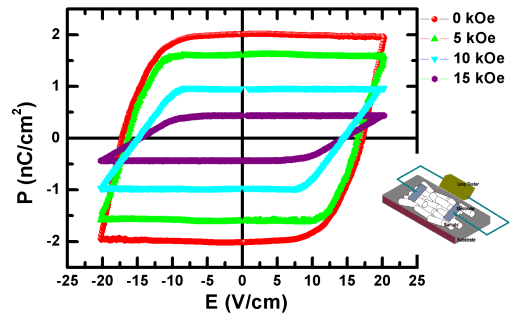}} 
   \subfigure[]{\includegraphics[scale=0.25]{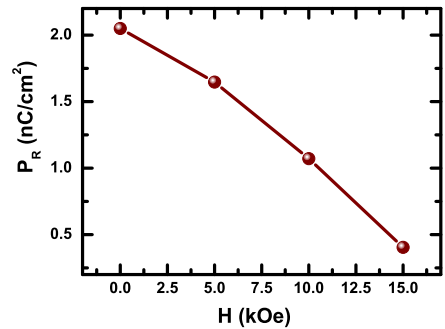}} 
   \end{center}
\caption{(a) The remanent hysteresis loops - measured at room temperature - under different magnetic fields; inset shows the schematic of the sample-electrode configuration used for the measurement of remanent hysteresis loops; (b) magnetic field dependence of the remanent polarization $P_R$ at room temperature. }
\end{figure}

We finally measured the remanent ferroelectric hysteresis loops at room temperature under different magnetic fields 0, 5, 10, and 15 kOe [Fig. 7(a)]. The loops have been measured by using an involved protocol which sends out fourteen voltage pulses to switch the domains and measure both the switchable and nonswitchable components of the polarization. Elimination of the contribution of nonswitchable component from the overall polarization yields the switchable remanent polarization. The details of the protocol and its underlying physics have been described in Ref. [28]. The loop shape is consistent with that observed by others for different ferroelectric compounds \cite{Scott}. Interestingly, the remanent polarization $P_R$ is found to decrease by $\sim$80\% under $\sim$15 kOe field at room temperature [Fig. 7(b)]. The Figure 7(a) inset shows the typical sample and electrode configuration. The decrease in $P_R$ under 0-15 kOe field indicates strong magnetoelectric coupling in nanorods of YNMO at room temperature. Such strong coupling is expected in cases where ferroelectricity originates from magnetism and the magnetic structure - including the magnetic anisotropy - changes under a magnetic field. Because of lower dimension at the surface, large Rashba spin-orbit coupling is expected \cite{Fert} which, in turn, could stabilize the DM exchange interaction across Mn$^{3+}$-O$^{2-}$-Mn$^{3+}$ among the other possible superexchange and double exchange interactions. Distinct magnetic structure at the surface has earlier been observed in other systems as well \cite{Langridge}. Since the surface ferroelectricity originates here from the magnetic structure, because of change in the magnetic anisotropy and/or structure under field, ferroelectric polarization too should change. It turns out that the change in the magnetic structure and/or switch in anisotropy (quantified by switch or rotation of the DM vector) under field in individual nanorod leads to decrease in overall polarization when summed up over the entire ensemble of nanorods studied. Of course, more detailed experimental as well as theoretical work needs to be done to unravel the surface magnetic structure and its change under such moderate magnetic field 0-15 kOe at room temperature. The interesting results presented in this paper on nanorods of double perovskite Y$_2$NiMnO$_6$ should trigger deeper investigation.  

\section{Summary}

In summary, we observed surface multiferroicity - magnetism, ferroelectricity, and significantly large magnetoelectric coupling - at room temperature in nanorods of double perovskite Y$_2$NiMnO$_6$ compound where bulk multiferroicity is observed only below $\sim$70 K. The surface magnetism has been probed by global magnetic measurements as well as imaging by magnetic force microscopy. It is found to be comprised of both ferromagnetic as well as antiferromagnetic domains. The Dzyloshinskii-Moriya exchange coupling interaction appears to stabilize and yield finite remanent ferroelectric polarization. Large magnetoelectric coupling, observed in this system, should trigger fresh research on other such candidate nanosized compounds for opening a new pathway of inducing room temperature surface multiferroicity even if its bulk counterpart either does not exist or exists only at low temperature. 

\begin{center}
$\textbf{ACKNOWLEDGMENTS}$
\end{center}

Two of the authors (S.M. and A.S.) acknowledge support (INSPIRE fellowship) from the Department of Science and Technology, Government of India, during this work.


\begin{thebibliography}   {99}
\bibitem{Fiebig} M. Fiebig, T. Lottermoser, D. Meier, and M. Trassin, Nat. Rev. Mater. $\textbf{1}$, 16046 (2016). 
\bibitem{Zutic} I. Zutic, A. Matos-Abiague, B. Scharf, H. Dery, and K. Belaschenko, Materials Today $\textbf{22}$, 85 (2019).
\bibitem{Spaldin} N.A. Spaldin and R. Ramesh, Nat. Mater. $\textbf{18}$, 203 (2019).   
\bibitem{Salje} J. Fontcuberta, V. Skumryev, V. Laukhin, X. Granados, and E.K.H. Salje, Scientific Reports $\textbf{5}$, 13784 (2015).
\bibitem{Bellaiche} Y. Yang, H. Xiang, H. Zhao, A. Stroppa, J. Zhang, S. Cao, J. Iniguez, L. Bellaiche, and W. Ren, Phys. Rev. B $\textbf{96}$, 104431 (2017). 
\bibitem{Mathur} See, for example, N. Mathur, M.J. Stolt, and S. Jin, APL Mater. $\textbf{7}$, 120703 (2019). 
\bibitem{Tokura} S. Seki, X.Z. Yu, S. Ishiwata, and Y. Tokura, Science $\textbf{336}$, 198 (2012). 
\bibitem{Loidl} E. Ruff, S. Widmann, P. Lunkenheimer, V. Tsurkan, S. Bordacs, I. Kezsmarki, and A. Loidl, Science Advances $\textbf{1}$, e1500916 (2015).
\bibitem{Kezsmarki} I. Kezsmarki, S. Bordacs, P. Milde, E. Neuber, L.M. Eng, J.S. White, H.M. Ronnow, C.D. Dewhurst, M. Mochizuki, K. Yanai $\textit{et al}$., Nat. Mater. $\textbf{14}$, 1116 (2015). 
\bibitem{Nahas} Y. Nahas, S. Prokhorenko, L. Louis, Z. Gui, I. Kornev, and L. Bellaiche, Nat. Commun. $\textbf{6}$, 8542 (2015).
\bibitem{Wang} L. Wang, Q. Feng, Y. Kim, R. Kim, K.H. Lee, S.D. Pollard, Y.J. Shin, H. Zhou, W. Peng, D. Lee $\textit{et al}$., Nat. Mater. $\textbf{17}$, 1087 (2018). 
\bibitem{Goncalves} M.A.P. Goncalves, C. Escorihuela-Sayalero, P. Garca-Fernandez, J. Junquera, and J. Iniguez, Science Advances $\textbf{5}$, eaau7023 (2019).
\bibitem{Das} S. Das, Y.L. Tang, Z. Hong, M.A.P. Goncalves, M.R. McCarter, C. Klewe, K.X. Nguyen, F. Gomez-Ortiz, P. Shafer, E. Arenholtz $\textit{et al}$., Nature (London) $\textbf{568}$, 368 (2019). 
\bibitem{Fert} See, for example, A. Soumyanarayanan, N. Reyren, A. Fert, and C. Panagopoulas, Nature (London) $\textbf{539}$, 509 (2016). 
\bibitem{Betouras} A.R. Tarkhany, M. Discacciati, and J.J. Betouras, Phys. Rev. B $\textbf{103}$, 205409 (2021).
\bibitem{Sundaresan} A. Sundaresan, R. Bhargavi, N. Rangarajan, U. Siddesh, and C.N.R. Rao, Phys. Rev. B $\textbf{74}$, 161306(R) (2006). 
\bibitem{Lu} C. Lu, W. Hu, Y. Tian, and T. Wu, Appl. Phys. Rev. $\textbf{2}$, 021304 (2015). 
\bibitem{Su} J. Su, Z.Z. Yang, X.M. Lu, J.T. Zhang, L. Gu, C.J. Lu, Q.C. Li, J.-M. Liu, and J.S. Zhu, ACS Appl. Mater. Interfaces $\textbf{7}$, 13260 (2015).
\bibitem{Supplemental} See the Supplemental Material for the structural details of the bulk and nanorods of Y$_2$NiMnO$_6$, high resolution transmission electron microscopy images and their analyses, x-ray photoelectron spectroscopy data for the bulk sample, results of Monte-Carlo simulation for different types of assembled nanorods, and the details of the analyses of magnetic force microscopy images. It is available on request from the corresponding authors. 
\bibitem{Zhao} See, for example, J. Wang, W. Wu, F. Zhao, and G.-M. Zhao, Appl. Phys. Lett. $\textbf{98}$, 083107 (2011). 
\bibitem{Usov} See, for example, N.A. Usov, O.N. Serebryakova, and V.P. Tarasov, Nanoscale Research Letters $\textbf{12}$, 489 (2017). 
\bibitem{Sahoo} A. Sahoo, D. Bhattacharya, and P.K. Mohanty, Phys. Rev. B $\textbf{101}$, 064414 (2020). 
\bibitem{Puntes} V.F. Puntes, P. Gorostiza, D.M. Aruguete, N.G. Bastus, and A.P. Alivisatos, Nat. Mater. $\textbf{3}$, 263 (2004).     
\bibitem{Aliyu} H.D. Aliyu, J.M. Alonso, P. de la Presa, W.E. Pottker, B. Ita, M. Garcia-Hernandez, and A. Hernando, Chem. Mater. $\textbf{30}$, 7138 (2018).
\bibitem{Nagaosa} See, for example, Y. Tokura, S. Seki, and N. Nagaosa, Rep. Prog. Phys. $\textbf{77}$, 076501 (2014). 
\bibitem{Gross} I. Gross, L.J. Martinez, J.-P. Tetienne, T. Hingant, J.-F. Roch, K. Garcia, R. Soucaille, J.P. Adam, J.-V. Kim, S. Rohart $\textit{et al}$., Phys. Rev. B $\textbf{94}$, 064413 (2016).
\bibitem{Chaurasiya} A.K. Chaurasiya, A. Kumar, R. Gupta, S. Chaudhary, P.K. Muduli, and A. Barman, Phys. Rev. B $\textbf{99}$, 035402 (2019).   
\bibitem{Chowdhury} U. Chowdhury, S. Goswami, D. Bhattacharya, A. Midya, and P. Mandal, Appl. Phys. Lett. $\textbf{109}$, 092902 (2016). 
\bibitem{Scott} J. Gardner and J.F. Scott, Materials Today $\textbf{21}$, 553 (2018). 
\bibitem{Langridge} S. Langridge, G.M. Watson, D. Gibbs, J.J. Betouras, N.I. Gidopoulos, F. Pollmann, M.W. Long, C. Vettier, and G.H. Lander, Phys. Rev. Lett. $\textbf{112}$, 167201 (2014).  
   
 
    

\end{thebibliography}
\end{document}